\RequirePackage{etex}
\documentclass[aps,pra, 10pt,twocolumn,superscriptaddress,floatfix]{revtex4-2}
\usepackage[english]{babel}
\usepackage[ascii]{inputenc}
\usepackage{amsmath,amssymb,amsfonts,amsthm}
\usepackage{mathtools}
\usepackage{braket}
\usepackage{graphicx}
\usepackage{float}
\usepackage{tikz-network}
\usepackage{tikz}
\usepackage{caption}
\usepackage{subcaption}
\usepackage{algorithm}
\usepackage{algpseudocode}
\usepackage{float}
\usepackage{makecell}
\usepackage[version=4]{mhchem}
\usepackage[colorlinks=true, bookmarks=false, allcolors=blue]{hyperref}
\usepackage{orcidlink}
\usepackage{quantikz}
\usepackage[compat=<version>]{yquant}
\usepackage{pifont}
\usepackage{multirow}
\usepackage{xcolor}
\usepackage{changes}

\usetikzlibrary{decorations.pathreplacing, decorations.text}

\DeclareUnicodeCharacter{00F6}{\"o}

\DeclarePairedDelimiter\abs{\lvert}{\rvert}
\DeclarePairedDelimiter\norm{\lVert}{\rVert}

\graphicspath{{figures/}}
\makeatletter
\def\input@path{{figures/}}
\makeatother

\begin{document}

\title{A Quantum Algorithm with Polylogarithmic Depth per Trotter Step for the Extended Hubbard Model}

\author{Yu Wang\,\orcidlink{0009-0004-3972-4388}}
\email{18yu.wang@tum.de}
\affiliation{Technical University of Munich, CIT, Department of Computer Science, Boltzmannstra{\ss}e 3, 85748 Garching, Germany}

\author{Martina Nibbi\,\orcidlink{0009-0001-6440-0498}}
\affiliation{Technical University of Munich, CIT, Department of Computer Science, Boltzmannstra{\ss}e 3, 85748 Garching, Germany}
\affiliation{Munich Center for Quantum Science and Technology, Schellingstra{\ss}e 4, 80799 M\"unchen, Germany}

\author{Maxine Luo\,\orcidlink{0009-0001-0513-3959}}
\affiliation{Munich Center for Quantum Science and Technology, Schellingstra{\ss}e 4, 80799 M\"unchen, Germany}
\affiliation{Max-Planck-Institute f\"ur Quantenoptik, Hans-Kopfermann-Stra{\ss}e 1, 85748 Garching, Germany}

\author{Isabel~Nha~Minh~Le\,\orcidlink{0000-0001-6707-044X}}
\affiliation{Technical University of Munich, CIT, Department of Computer Science, Boltzmannstra{\ss}e 3, 85748 Garching, Germany}
\affiliation{Munich Center for Quantum Science and Technology, Schellingstra{\ss}e 4, 80799 M\"unchen, Germany}

\author{Yanbin Chen\,\orcidlink{0000-0002-1123-1432}}
\affiliation{Technical University of Munich, CIT, Department of Computer Science, Boltzmannstra{\ss}e 3, 85748 Garching, Germany}

\author{J.~Ignacio Cirac\,\orcidlink{0000-0003-3359-1743}}
\affiliation{Munich Center for Quantum Science and Technology, Schellingstra{\ss}e 4, 80799 M\"unchen, Germany}
\affiliation{Max-Planck-Institute f\"ur Quantenoptik, Hans-Kopfermann-Stra{\ss}e 1, 85748 Garching, Germany}

\author{Christian B.~Mendl\,\orcidlink{0000-0002-6386-0230}}
\affiliation{Technical University of Munich, CIT, Department of Computer Science, Boltzmannstra{\ss}e 3, 85748 Garching, Germany}
\affiliation{Technical University of Munich, Institute for Advanced Study, Lichtenbergstra{\ss}e 2a, 85748 Garching, Germany}

\date{\today}

\begin{abstract}

The extended Hubbard model on a two-dimensional lattice captures key physical phenomena, but its simulation remains challenging because long-range interactions give rise to a large number of interaction terms. Here we present Q2FMM, an efficient quantum algorithm for simulating this model within the Trotter product formula. Inspired by the fast multipole method, Q2FMM replaces site-site interactions with interactions between hierarchical coarse-grained boxes across multiple length scales. In addition, the multipole expansions of boxes are reused for their parent boxes, further enhancing the efficiency. To enable this hierarchical reuse coherently, we design a reversible quantum circuit that removes garbage information through uncomputing. The resulting circuit depth for a single Trotter step scales polylogarithmically with system size.

\end{abstract}

\maketitle

Quantum simulation of strongly correlated electronic structure and condensed matter problems is widely regarded as one of the most promising applications of quantum computing~\cite{feynman1982, lloyd1996universal, McArdle2020RMP, Georgescu2014RMP, reiher2017elucidating, arute2020hartreefock, preskill2018nisq, lin2022heisenberg, monroe2021programmable}. Within condensed-matter physics, the classical simulation of a two-dimensional (2D) system is a central challenge~\cite{PhysRevLett.125.100503, PRXQuantum.6.010312, RevModPhys.93.045003, PhysRevB.106.245102}, making them a natural target for quantum simulation.
Another particularly important challenge arises when including long-range interactions, which occur widely across physically important settings and can qualitatively alter the behavior of the system~\cite{defenu2023long, schuler2013optimal, wang2022experimental,landig2016quantum}. By the all-to-all couplings, they greatly increase the number of interaction terms and reduce the scope for parallelization because many terms overlap. Consequently, even for models defined on low-dimensional lattices, simulating the time evolution can require very deep quantum circuits. 
A paradigmatic example is the extended Fermi-Hubbard model~\cite{Hubbard1963, Arovas2022} with long-range (Coulomb) interactions on a 2D lattice, which plays a crucial role in investigating a variety of physical phenomena, including superconducting properties~\cite{Scalapino2012, jiang2018dwave, reymbaut2016antagonistic}, charge density waves~\cite{zhangextended1989, terletska2017extended}, and Mott insulator-metal transitions~\cite{Imada1998, Lee2006}. Moreover, it can be specialized for chemical systems, resulting in the Pariser-Parr-Pople (PPP) model~\cite{PariserParr1953b, Pople1953, Heeger1988}. These features make the extended Hubbard model a representative and physically relevant setting for studying quantum simulation in the presence of long-range interactions. Its Hamiltonian is given by:
\begin{equation}
\label{eq:hubbard}
\hspace{-0.2cm}
H = \underbrace{h \sum_{\substack{\langle a,b \rangle\\ \sigma}} \hat{c}_{a, \sigma}^\dagger \hat{c}_{b, \sigma}}_{T}
+ \underbrace{V_0 \sum_{\substack{a\\\,}} \hat{n}_{a\uparrow} \hat{n}_{a\downarrow}}_{V_{\text{os}}}
+ \underbrace{\frac{1}{2} \sum_{\substack{a \ne b\\ \sigma,\sigma'}} V_{ab} \hat{n}_{a\sigma} \hat{n}_{b\sigma'}}_{V_\text{C}},
\end{equation}
where $a,b$ enumerate the lattice sites and $\sigma, \sigma'\in\{\downarrow,\uparrow\}$ denote the spins.
The nearest-neighbor hopping is described by $T$, where $\hat{c}^{\dagger}_{a,\sigma}$ and $\hat{c}_{a,\sigma}$ are fermionic creation and annihilation operators.
$V_{\text{os}}$ models the on-site interaction, with $\hat{n}_{a,\sigma}$ denoting the number operator, and the long-range (Coulomb) interaction is captured by $V_\text{C}$.

When simulating the time evolution of the extended Hubbard model with long-range interactions on a quantum computer, the local term $V_{\text{os}}$ can be implemented using single-qubit gates in parallel. It is also well known that the hopping term $T$ on a 2D lattice can be realized with constant quantum circuit depth for a single Trotter step by employing local encoding methods, such as the Verstraete-Cirac encoding~\cite{verstraete2005mapping}, the Bravyi-Kitaev superfast encoding~\cite{bravyi2002fermionic}, and compact encoding~\cite{derby2021compact, clinton2021hamiltonian}. In contrast, simulating the long-range Coulomb term is more costly due to the $\mathcal{O}(N^2)$ interaction terms, where $N$ is the number of lattice sites. When employing the fermionic SWAP network method~\cite{kivlichan2018quantum}, the depth per Trotter step remains $\mathcal{O}(N)$. Lowering the cost is possible when long-range interactions decay rapidly with distance, as the interaction range can then be truncated so that only a limited number of interaction pairs are retained in the simulation~\cite{tran2019locality, Childs2021PRXCommutator}. However, this truncation strategy becomes less effective for slowly decaying interactions such as the Coulomb potential, which scales only as the inverse of the distance, since a large cutoff radius is required to maintain the desired accuracy.

In this work, we seek to coarse-grain the long-range Coulomb interaction by replacing site-site pairwise interactions with interactions between well-separated boxes. Rather than treating all pairwise terms individually, the simulation is reformulated in terms of local collective contributions and their interactions across multiple length scales. This naturally leads to a hierarchical description, for which the \emph{fast multipole method (FMM)}~\cite{barnes1986hierarchical, ying2012pedestrian, greengard1987fast, darve2000fast} provides a well-suited framework. Based on this idea, we introduce Q2FMM, an algorithm for simulating the time evolution of the second-quantized extended Fermi-Hubbard model, which employs FMM for systematically coarse-graining the Coulomb potential and reusing the intermediate computing results. We also discuss how the \texttt{COPY} operation~\cite{Nielsen_2011, gottesman2009, Sahay_2025, dellachiara2025lcu} and unbounded fan-out gates~\cite{hoyer2005quantum} can further improve overall efficiency.

The resulting quantum circuit of Q2FMM implements the time evolution of the long-range Coulomb term $V_\text{C}$ with polylogarithmic circuit depth per Trotter step. Moreover, the per-step gate complexity scales quasi-linearly with system size and polylogarithmically with target FMM error and finite-precision arithmetic error. Accounting for the Trotter steps required to simulate total time $t$ with target error $\epsilon$, the total depth and gate complexity scale as $\Tilde{O}\left(N^{1/2}t^{3/2}/\epsilon^{1/2}\right)$ and $\Tilde{O}\left(N^{3/2}t^{3/2}/\epsilon^{1/2}\right)$, respectively. Here, all error contributions are assumed to be of the same order and are collectively denoted by $\epsilon$, and the $\tilde{\mathcal{O}}$ notation indicates that additional polylogarithmic factors are not explicitly reported. While the Q2FMM framework is general, it may be particularly well suited for implementation on two-dimensional neutral-atom quantum computers~\cite{hollerith2022distance, tao2024highfidelity, wei2023braneparity, graham2022multi, evered2023high, bluvstein2022processor, schmid2024compiler, bluvstein2024logical, evered2025kitaev, manetsch2025tweezer, chiu2025continuous3000, xu2025neutralhubbard}, which naturally support atom shuttling~\cite{graham2022multi, evered2023high, bluvstein2022processor, schmid2024compiler, bluvstein2024logical}. It may also be implemented in a surface-code setting on a two-dimensional lattice, where long-range operations and fan-out gates can be realized in constant depth~\cite{litinski2019game, edp2022Beverland}.

\section{Results}

\subsection{From site-site interactions to box-box interactions}

For simplicity, we neglect the spin, which will be reconsidered in Appendix~\ref{app:reconsider}. In this case, each site can only contain at most one fermion. Also, we focus on a square lattice, but extending to other geometries (e.g., rectangular lattices) is straightforward. Furthermore, we specifically consider the Coulomb interaction described by the kernel
\begin{equation}
\label{eq:cou_kernel}
K_{\text{C}}(\mathbf{r}_a,\mathbf{r}_b) \coloneqq 1/\|\mathbf{r}_a - \mathbf{r}_b\|_2,
\end{equation}
where $\|\cdot\|_2$ denotes the Euclidean norm, $\mathbf{r}_a$ and $\mathbf{r}_b$ denote the position of sites $a$ and $b$, respectively. Hence, we focus on simulating:
\begin{equation}
V_C =
\frac{1}{2}\sum_{a\ne b}
K_C(\mathbf r_a,\mathbf r_b)\hat n_a \hat n_b ,
\end{equation}
as the on-site term $V_{\text{os}}$ and kinetic term $T$ can be simulated efficiently, as discussed.

A natural strategy for simplifying the long-range Coulomb interaction is to coarse-grain it, replacing the original site-site pairwise interactions by interactions between well-separated regions, referred to as ``boxes'' in this work. The simplest approximation is to ignore the internal variation within a distant box and represent its effect by a single interaction value. Concretely, for two well-separated boxes $A$ and $B$ containing points $a\in A$ and $b\in B$, the pairwise kernel $K_{\text{C}}(\mathbf{r}_a,\mathbf{r}_b)$ is approximated by the representative value $K_{\text{C}}(\mathbf{r}_A,\mathbf{r}_B)$:
\begin{equation}
V_{AB} \coloneqq \frac{1}{2} \sum_{\substack{a\in A\\b\in B}} K_{\text{C}}(\mathbf{r}_a,\mathbf{r}_b)\hat{n}_a\hat{n}_b 
\approx \frac{1}{2}K_{\text{C}}(\mathbf{r}_A,\mathbf{r}_B)\hat{N}_A\hat{N}_B,
\label{eq:HAB}
\end{equation}
where $\mathbf{r}_A$ and $\mathbf{r}_B$ denote the centers of the two boxes (see Fig.~\ref{fig:point1}), and $\hat{N}_A = \sum_{a\in A}\hat{n}_a$. In this way, the simulation is reformulated from resolving individual site-to-site interactions to capturing the collective interaction between distant boxes.

\begin{figure*}
\centering
\makebox[\textwidth][c]{\hspace*{-1.5cm}\begin{tikzpicture}[scale=0.6]

  \draw[step=0.5cm, line width=0.5pt, color=blue!30!black] (-7,0) grid (-1,6);

  \filldraw[fill=yellow, fill opacity=0.4, draw=black]
    (-5.5,1.5) rectangle (-5,2);
  \filldraw[fill=yellow, fill opacity=0.4, draw=black]
    (-5,1.5) rectangle (-4.5,2);
  \filldraw[fill=yellow, fill opacity=0.4, draw=black]
    (-4.5,1.5) rectangle (-4,2);

  \filldraw[fill=yellow, fill opacity=0.4, draw=black]
    (-5.5,2) rectangle (-5,2.5);
  \filldraw[fill=yellow, fill opacity=0.4, draw=black]
    (-4.5,2) rectangle (-4,2.5);

  \filldraw[fill=yellow, fill opacity=0.4, draw=black]
    (-5.5,2.5) rectangle (-5,3);
  \filldraw[fill=yellow, fill opacity=0.4, draw=black]
    (-5,2.5) rectangle (-4.5,3);
  \filldraw[fill=yellow, fill opacity=0.4, draw=black]
    (-4.5,2.5) rectangle (-4,3);

  \filldraw[fill=red, fill opacity=0.4, draw=black]
    (-5,2) rectangle (-4.5,2.5);
  \node[scale=0.8] at (-4.75,2.25) {$A$};

  \node at (-4,-0.35) {(a)};

  \draw[step=0.5cm, dashed, line width=0.5pt, color=blue!30!black] (0,0) grid (6,6);

  \filldraw[fill=black, fill opacity=0.4, draw=black, even odd rule]
    (0,0) rectangle (6,6)
    (1,1) rectangle (4,4);

  \draw[step=1cm, line width=1pt, color=green!40!black] (0,0) grid (6,6);

  \filldraw[fill=red, fill opacity=0.4, draw=black]
    (2,2) rectangle (2.5,2.5);
  \node[scale=0.8] at (2.25,2.25) {$A$};

  \node at (3,-0.35) {(b)};

  \draw[step=0.5cm, line width=0.5pt, color=blue!30!black] (7,0) grid (13,6);

  \filldraw[fill=black, fill opacity=0.4, draw=black, even odd rule]
    (7,0) rectangle (13,6)
    (8,1) rectangle (11,4);

  \filldraw[fill=blue!70, fill opacity=0.4, draw=black, even odd rule]
    (8,1) rectangle (11,4)
    (8.5,1.5) rectangle (10,3);

  \filldraw[fill=yellow, fill opacity=0.4, draw=black]
    (8.5,1.5) rectangle (9,2);
  \filldraw[fill=yellow, fill opacity=0.4, draw=black]
    (9,1.5) rectangle (9.5,2);
  \filldraw[fill=yellow, fill opacity=0.4, draw=black]
    (9.5,1.5) rectangle (10,2);

  \filldraw[fill=yellow, fill opacity=0.4, draw=black]
    (8.5,2) rectangle (9,2.5);
  \filldraw[fill=yellow, fill opacity=0.4, draw=black]
    (9.5,2) rectangle (10,2.5);

  \filldraw[fill=yellow, fill opacity=0.4, draw=black]
    (8.5,2.5) rectangle (9,3);
  \filldraw[fill=yellow, fill opacity=0.4, draw=black]
    (9,2.5) rectangle (9.5,3);
  \filldraw[fill=yellow, fill opacity=0.4, draw=black]
    (9.5,2.5) rectangle (10,3);

  \filldraw[fill=red, fill opacity=0.4, draw=black]
    (9,2) rectangle (9.5,2.5);
  \node[scale=0.8] at (9.25,2.25) {$A$};

  \node[scale=0.8] at (10.75,3.75) {$B$};

  \draw[<->, >=stealth', shorten <=8pt, shorten >=8pt, red!50!black, line width=1pt]
    (9.25,2.25) -- (10.75,3.75);

  \node at (10,-0.35) {(c)};

  \draw[step=0.5cm, dashed, line width=0.5pt, color=blue!30!black] (-7,-7) grid (-1,-1);

  \draw[step=1cm, line width=1pt, color=green!40!black] (-7,-7) grid (-1,-1);

  \draw[->, >=stealth', shorten <=4pt, shorten >=4pt, red!50!black, line width=1pt]
    (-6.5,-6.5) -- (-7.55,-7.05);
  \node[draw=black, rounded corners=1pt, inner sep=1pt, font=\scriptsize] at (-7.72,-7.24) {adder};

  \node at (-4,-7.35) {(d)};

  \draw[step=1cm, line width=1pt, color=black] (0,-7) grid (6,-1);

  \filldraw[fill=black, fill opacity=0.4, draw=black]
    (4,-7) rectangle (6,-1);

  \filldraw[fill=blue!70, fill opacity=0.4, draw=black, even odd rule]
    (0,-7) rectangle (4,-1)
    (0,-6) rectangle (3,-3);

  \filldraw[fill=yellow, fill opacity=0.4, draw=black]
    (0,-6) rectangle (1,-5);
  \filldraw[fill=yellow, fill opacity=0.4, draw=black]
    (1,-6) rectangle (2,-5);
  \filldraw[fill=yellow, fill opacity=0.4, draw=black]
    (2,-6) rectangle (3,-5);

  \filldraw[fill=yellow, fill opacity=0.4, draw=black]
    (0,-5) rectangle (1,-4);
  \filldraw[fill=yellow, fill opacity=0.4, draw=black]
    (2,-5) rectangle (3,-4);

  \filldraw[fill=yellow, fill opacity=0.4, draw=black]
    (0,-4) rectangle (1,-3);
  \filldraw[fill=yellow, fill opacity=0.4, draw=black]
    (1,-4) rectangle (2,-3);
  \filldraw[fill=yellow, fill opacity=0.4, draw=black]
    (2,-4) rectangle (3,-3);

  \filldraw[fill=red, fill opacity=0.4, draw=black]
    (1,-5) rectangle (2,-4);

  \node at (1.5,-4.5) {$A^{\prime}$};

  \node at (3,-7.35) {(e)};

  \filldraw[fill=blue!70, fill opacity=0.4, draw=black] (11,-7) rectangle (13,-5);
  \filldraw[fill=blue!70, fill opacity=0.4, draw=black] (11,-5) rectangle (13,-3);
  \filldraw[fill=blue!70, fill opacity=0.4, draw=black] (7,-3) rectangle (9,-1);
  \filldraw[fill=blue!70, fill opacity=0.4, draw=black] (9,-3) rectangle (11,-1);
  \filldraw[fill=blue!70, fill opacity=0.4, draw=black] (11,-3) rectangle (13,-1);

  \filldraw[fill=yellow, fill opacity=0.4, draw=black] (9,-7) rectangle (11,-5);
  \filldraw[fill=yellow, fill opacity=0.4, draw=black] (7,-5) rectangle (9,-3);
  \filldraw[fill=yellow, fill opacity=0.4, draw=black] (9,-5) rectangle (11,-3);

  \filldraw[fill=red, fill opacity=0.4, draw=black] (7,-7) rectangle (9,-5);

  \node[scale=1.0] at (8,-6) {$A^{\prime\prime}$};

  \node at (10,-7.35) {(f)};

\end{tikzpicture}}
\caption{(a) The near field of a target box $A$ (red) is defined by its neighboring boxes, shown in yellow. (b) The far field (black) consists of boxes that are not neighboring boxes of the target box's parent box. The corresponding parent boxes are circled in green. (c) Boxes not in the near or far field form the interaction list $I(A)$ (yellow) at an arbitrary level $L$. For $B \in I(A)$ and $a \in A$, $b \in B$, we approximate $K_{\text{C}}(\mathbf{r}_a,\mathbf{r}_b)$ by $K_{\text{C}}(\mathbf{r}_A,\mathbf{r}_B)$ in the $0^\text{th}$-order FMM. (d) At a coarser level $L-1$, the occupation number of a parent box is obtained by directly summing the occupation numbers of its child boxes, which can be realized by quantum adders. (e) Example of the interaction list of a box at a coarser level. The same rule for constructing the interaction list applies at every level. (f) As one moves to an even coarser level $L-2$, four child boxes are merged into one parent box, and the interaction list of a given box is presented. This procedure is repeated until the coarsest level is reached. The implementation of time evolution is not drawn for visual clarity.}
\label{fig:point1}
\end{figure*}

We make two comments on this approximation. First, it is accurate when the two boxes are sufficiently far apart, and each box is sufficiently small. As the boxes approach each other, however, the approximation becomes less accurate, since it ignores the internal structure of each box. Second, the admissible box size naturally depends on the separation between the two boxes: larger separations permit larger boxes, whereas shorter separations require finer partitions. This suggests that the interaction should be organized systematically according to both distance and box size. These observations motivate a hierarchical multiscale representation that adapts the box size to the separation between boxes and systematically incorporates the internal structure of each box when needed. This is exactly the role of the fast multipole method (FMM)~\cite{barnes1986hierarchical, ying2012pedestrian, greengard1987fast, darve2000fast}, which accounts for the electron distribution and organizes boxes hierarchically across different length scales. From this viewpoint, hierarchically organizing the simple approximation discussed above corresponds precisely to the $0^\text{th}$-order FMM.

\subsection{Time evolution via the $0^\text{th}$-order Q2FMM}
\label{sec:0th_q2fmm}

To convey the core idea of our quantum algorithm, we begin with a simplified construction based on the $0^\text{th}$-order FMM, which we refer to as the $0^\text{th}$-order Q2FMM. This simplified formulation highlights the essential ingredients of the algorithm and provides a convenient setting for introducing the hierarchical boxes in the quadtree structure, as well as the notions of near-field and far-field interactions. The higher-order FMM is presented later in Sec.~\ref{sec:higher_FMM}, where the internal structure of each box is taken into account, thereby enabling a systematic improvement in the accuracy of the algorithm.

Fig.~\ref{fig:point1} illustrates the $0^\text{th}$-order Q2FMM in terms of a hierarchy of boxes organized as a quadtree. This organization offers two main advantages. First, it makes both the interaction distances and the corresponding box sizes explicit. Second, it allows information from smaller boxes to be systematically aggregated and reused at the level of larger boxes. The higher-order Q2FMM also follows such a structure.

In such a quadtree structure, level $L$ contains $4^{L}$ boxes for a square lattice, and the finest level is denoted by $L_{\max}=\log_4 N$, assuming a square lattice. Our algorithm proceeds from the finest level $L = L_{\max}$ down to the second-coarsest level $L=2$, which contains $4^2$ boxes. The \emph{parent box} of a given box is the next larger box that directly contains it, while its \emph{child boxes} are the smaller boxes obtained by subdividing it at its finer level. Equivalently, the parent of a box at level $L+1$ is the box at level $L$ that contains it. We present how to determine the interaction distances as follows. As detailed in Fig.~\ref{fig:point1}, for each box, all other boxes are divided into three categories at an arbitrary level in the quadtree based on their distance from the current box: neighbor boxes form the \emph{near field}, boxes not in the near field but whose parents are neighbors of the current box's parent form the \emph{interaction list}, and the remaining ones belong to the \emph{far field}. At each level, the summation of point-point interactions between box $A$ and box $B$ is approximated by Eq.~\eqref{eq:HAB}. For each box, we only consider interactions with boxes within its interaction list that are sufficiently far away to ensure the accuracy of the approximation, while avoiding double-counting of interactions that have already been handled at coarser levels.
Consequently, for each level $L$, which contains $4^{L}$ boxes in total, we only calculate the box-box interactions in each box $A$'s interaction list $I(A)$:
\begin{equation}
    V_L \coloneqq
    \frac{1}{2} \sum_A \sum_{B \in I_L(A)} K_{\mathrm{C}}(\mathbf{r}_A,\mathbf{r}_B)\hat N_A \hat N_B.
\label{eq:Hl}
\end{equation}
The finest level $L=L_{\max}$ is treated as a special case: in this case, $V_L$ also  collects the interaction pairs lying within the near field at the finest level.
Thus, we can approximate the Coulomb term as:
\begin{equation}
    V_\text{C} = \frac{1}{2} \sum_{a \ne b} K_{\mathrm{C}}(\mathbf{r}_a,\mathbf{r}_b) \hat{n}_{a} \hat{n}_{b} \approx  \sum_{L=2}^{L_{\text{max}}} V_{L},
\label{eq:H_app}
\end{equation}
where $L_{\max}$ denotes the finest level of the hierarchy.
As illustrated in Fig.~\ref{fig:defer}, all interaction pairs are accounted for by systematically traversing all levels of the hierarchy in this manner.

Following the Trotterization technique~\cite{Trotter1959Product, Suzuki1976Generalized, Berry2007Sparse, Childs2019LatticePF, Childs2021PRXCommutator}, we can approximate the time evolution of $H$ using the second-order Trotterization:
\begin{equation}
\begin{aligned}
\label{eq:trotter}
     U(t) &\coloneqq e^{-i t (T+V_\text{os}+V_\text{C})} \\
     &\approx \left( e^{-i\frac{\delta t}{2}T} \, e^{-i\delta t (V_\text{C} + V_\text{os})} \, e^{-i\frac{\delta t}{2}T} \right)^d \\
     &= \left( e^{-i\frac{\delta t}{2}T} \, e^{-i\delta t V_\text{C}}\, e^{-i\delta t V_\text{os}} \, e^{-i\frac{\delta t}{2}T} \right)^d,
\end{aligned}
\end{equation}
where $d$ is the number of Trotter steps and $\delta t = t/d$. In this work, we focus on the time evolution operator $e^{-i\delta t V_\text{C}}$ that corresponds to the Coulomb term, for which we employ the approximation from Eq.~\eqref{eq:H_app}, so that
\begin{equation}
     e^{-i\delta t V_\text{C}} \approx  \prod_{L=2}^{L_{\max}} e^{-i \delta t V_{L}}. 
\label{eq:time_total}
\end{equation}
By inserting Eq.~\eqref{eq:Hl}, we get
\begin{equation}
\label{eq:time_l}
    e^{-i \delta t V_{L}} =  \prod_{A} \prod_{B\in {I_{L}(A)}} e^{-i t^{\prime}_{AB} \hat{N}_A\hat{N}_B},
\end{equation}
where we absorbed all the coefficients into the effective time
\begin{equation}
t^{\prime}_{AB} = K_{\text{C}}(\mathbf{r}_A,\mathbf{r}_B)\delta t/ 2.
\end{equation}
In Eq.~\eqref{eq:time_l}, $I_L(A)$ denotes the interaction list associated with level $L$ and the boxes with corresponding size. 

\begin{figure}
\centering
\begin{tikzpicture}[scale=0.6]

  \draw[step=0.5cm, line width=0.5pt, color=blue!30!black] (0,0) grid (6,6);

  \filldraw[fill=black, fill opacity=0.4, draw=black, even odd rule]
    (0,0) rectangle (6,6)
    (1,1) rectangle (4,4);

  \filldraw[fill=blue!70, fill opacity=0.4, draw=black, even odd rule]
    (1,1) rectangle (4,4)
    (1.5,1.5) rectangle (3,3);

  \filldraw[fill=yellow, fill opacity=0.4, draw=black]
    (1.5,1.5) rectangle (2,2);
  \filldraw[fill=yellow, fill opacity=0.4, draw=black]
    (2,1.5) rectangle (2.5,2);
  \filldraw[fill=yellow, fill opacity=0.4, draw=black]
    (2.5,1.5) rectangle (3,2);

  \filldraw[fill=yellow, fill opacity=0.4, draw=black]
    (1.5,2) rectangle (2,2.5);
  \filldraw[fill=yellow, fill opacity=0.4, draw=black]
    (2.5,2) rectangle (3,2.5);

  \filldraw[fill=yellow, fill opacity=0.4, draw=black]
    (1.5,2.5) rectangle (2,3);
  \filldraw[fill=yellow, fill opacity=0.4, draw=black]
    (2,2.5) rectangle (2.5,3);
  \filldraw[fill=yellow, fill opacity=0.4, draw=black]
    (2.5,2.5) rectangle (3,3);

  \filldraw[fill=red, fill opacity=0.4, draw=black]
    (2,2) rectangle (2.5,2.5);
  \node[scale=0.8] at (2.25,2.25) {$A$};

  \node[scale=0.55] at (3.25,3.75) {$C_1$};
  \node[scale=0.55] at (3.75,3.75) {$C_2$};
  \node[scale=0.55] at (3.25,3.25) {$C_3$};
  \node[scale=0.55] at (3.75,3.25) {$C_4$};

  \node at (3,-0.35) {(a)};

  \begin{scope}[xshift=8cm]
    \draw[step=1cm, line width=1pt, color=black] (0,0) grid (6,6);

    \filldraw[fill=blue!70, fill opacity=0.4, draw=black, even odd rule]
      (0,0) rectangle (6,6)
      (1,1) rectangle (4,4);

    \filldraw[fill=yellow, fill opacity=0.4, draw=black]
      (1,1) rectangle (2,2);
    \filldraw[fill=yellow, fill opacity=0.4, draw=black]
      (2,1) rectangle (3,2);
    \filldraw[fill=yellow, fill opacity=0.4, draw=black]
      (3,1) rectangle (4,2);

    \filldraw[fill=yellow, fill opacity=0.4, draw=black]
      (1,2) rectangle (2,3);
    \filldraw[fill=yellow, fill opacity=0.4, draw=black]
      (3,2) rectangle (4,3);

    \filldraw[fill=yellow, fill opacity=0.4, draw=black]
      (1,3) rectangle (2,4);
    \filldraw[fill=yellow, fill opacity=0.4, draw=black]
      (2,3) rectangle (3,4);
    \filldraw[fill=yellow, fill opacity=0.4, draw=black]
      (3,3) rectangle (4,4);

    \filldraw[fill=red, fill opacity=0.4, draw=black]
      (2,2) rectangle (3,3);

    \node at (2.5,2.5) {$A^{\prime}$};

    \node at (3.5,3.5) {$C$};

    \node at (3,-0.35) {(b)};
  \end{scope}

\end{tikzpicture}
\caption{The interaction between $A'$ and $C$, which is not evaluated at the current level, is instead accounted for at finer levels, in the form of interactions between the child box $A$ and the boxes $C_1$, $C_2$, $C_3$, and $C_4$. Although the present example only shows one child box of $A'$, the full interaction between $A'$ and $C$ is recovered when the same procedure is applied to all child boxes throughout the complete workflow.}
\label{fig:defer}
\end{figure}

While it is straightforward to apply the FMM in classical computation to evaluate the total interaction, realizing it on a quantum computer for time evolution is nontrivial. There are two concerns. First, if one simply applies time-evolution operators $e^{-i t^{\prime}_{AB} \hat{N}_A \hat{N}_B}$ to pairs of boxes, then the correct dynamical phase is generated, but at a high computational cost. At finer levels, this cost arises from the large number of box-box interaction terms, whereas at coarser levels it arises from the expensive computation of the box-box interaction, since each box contains many grid points. Thus, the reasonable reuse of intermediate results needs to be proposed. Second, even if the information from smaller boxes is aggregated into larger ones, e.g., by summing their occupation numbers as discussed later, such a procedure generally introduces additional entanglement or leaves behind garbage information. In this work, we design a quantum circuit that realizes the hierarchical information flow underlying the FMM while enabling efficient, coherent reuse of information without leaving behind extra entanglement or garbage states.

\subsection{Circuit design for $0^\text{th}$-order Q2FMM}
\label{sec:0thcircuit}

In this subsection, we propose a quantum circuit that can address the concerns discussed above.
We start from the observation that the evolution of each computational basis under $e^{-i \delta t V_L}$ depends on the product $N_A N_B$, where $N_A$ and $N_B$ are the respective occupation numbers of box $A$ and box $B$. 
An efficient way to simulate  $e^{-i \delta t V_L}$ consists of the following steps:
\begin{enumerate}
    \item Calculate the occupation numbers $N_A$ and $N_B$ for boxes $A$ and $B$ separately.
    \item Compute the product $N_A N_B$. 
    \item Compute the evolving phase $e^{- i t^{\prime}_{AB} {N}_A {N}_B}$.
\end{enumerate}
At the finest level of the hierarchy, $L = L_{\text{max}} = \log_4(N)$, the boxes coincide with the physical two-dimensional lattice. 
At this resolution, each lattice site corresponds to a single qubit, where $\ket{1}$ denotes an occupied site and $\ket{0}$ an empty one. Also, the qubit layout directly corresponds to the positions of the lattice sites; hence, the qubits that need to be accessed in order to evaluate the Coulomb interactions are already known in advance. The time evolution for the finest level is trivially the direct implementation of local terms $e^{-i\delta t \hat{n}_a\hat{n}_b}$.
Proceeding to the next level, the occupation number of each box is obtained as the arithmetic sum of the four qubits contained within it. 
For all coarser levels, the occupation number of a box is recursively determined by summing the occupation numbers of its child boxes, as illustrated in Fig.~\ref{fig:point1}. Consequently, all parent boxes at coarser levels can reuse the occupation information from their child boxes at the previous level. This summation can be carried out using a quantum adder~\cite{wang2025comprehensive, draper2006logarithmic, takahashi2009quantum, draper2000addition, remaud2025ancilla}, as exemplified in Fig.~\ref{fig:adder}. 
Later, we will further analyze the computational complexity of this procedure and its contribution to the overall cost of the algorithm.

\begin{figure}
\centering
\begin{subfigure}{0.45\textwidth}
    \centering
    \begin{quantikz}[row sep=0.1cm, column sep=0.4cm]
  \lstick{$\ket{a}$} & \qwbundle{n}
    & \gate[wires=3]{\text{\parbox{2.1cm}{\centering Quantum\\Adder}}}
    & \rstick{$\ket{a}$} \\
  \lstick{$\ket{b}$} & \qwbundle{n}
    & \ghost{\text{\parbox{2.1cm}{\centering Quantum\\Adder}}}
    & \rstick{$\ket{b}$} \\
  \lstick{$\ket{0}$} & \qwbundle{n+1}
    & \ghost{\text{\parbox{2.1cm}{\centering Quantum\\Adder}}}
    & \rstick{$\ket{a + b}$}
\end{quantikz}
    \caption{
    An out-of-place quantum adder, in which the output $\ket{a+b}$ is recorded in ancilla qubits.}
    \label{fig:adder}
\end{subfigure}

\begin{subfigure}{0.45\textwidth}
    \centering
    \begin{quantikz}[row sep=0.1cm, column sep=0.4cm]
  \lstick{$\ket{a}$} & \qwbundle{n}
    & \gate[wires=3]{\text{\parbox{2.1cm}{\centering Quantum\\Multiplier}}}
    & \rstick{$\ket{a}$} \\
  \lstick{$\ket{b}$} & \qwbundle{n}
    & \ghost{\text{\parbox{2.1cm}{\centering Quantum\\Multiplier}}}
    & \rstick{$\ket{b}$} \\
  \lstick{$\ket{0}$} & \qwbundle{2n}
    & \ghost{\text{\parbox{2.1cm}{\centering Quantum\\Multiplier}}}
    & \rstick{$\ket{a \cdot b}$}
\end{quantikz}
    \caption{An $n$-bit quantum multiplication gate. The binary representation of the result $\ket{a \cdot b}$ is stored in an ancilla register. }
    \label{fig:multiplier}
\end{subfigure}

\caption{Conceptual illustration of out-of-place quantum adders and multipliers. In contrast, in-place designs overwrite the input registers with the results. The explicit circuit structure varies across different implementations.}
\end{figure}

To perform step 2, we employ a quantum multiplier~\cite{wang2025comprehensive, kahanamoku2024fast, draper2006logarithmic, dutta2018quantum, kepley2015quantum, orts2023improving}, as exemplified in Fig.~\ref{fig:multiplier}.
The product $N_A N_B$ is encoded as a binary string on a \emph{product register} $\ket{N_A N_B}$.
Finally, in step 3, the evolving phase $e^{- i t^{\prime}_{AB} {N}_A {N}_B}$ is implemented by applying single-qubit phase gates of the form
\begin{equation}
\label{eq:phase}
    \begin{split}
    P_b &= \ket{0}\bra{0}+e^{-it^{\prime}_{AB} \cdot 2^{b}}\ket{1}\bra{1} \\
    &= \begin{pmatrix}
        1 & 0 \\
        0 & e^{-i t^{\prime}_{AB} \cdot 2^b}
    \end{pmatrix}
    \end{split}
\end{equation}
to each qubit $b$ of the product register $\ket{N_A N_B}$.
The cumulative action of the gates
\begin{equation}
\label{eq:sequence_p}
    U_{AB} = \prod_{b=0}^{n-1} P_b
\end{equation}
applies exactly the overall phase $e^{-i t^{\prime}_{AB} N_A N_B}$. Note that the gates $P_b$ can be applied in parallel. The obtained phase $e^{-i t^{\prime}_{AB} N_A N_B}$ is first imprinted on the ancilla qubits that record the product $N_A N_B$. We must uncompute the quantum multiplier to transfer the phase back to the qubits that represent the quantum state.

To reuse already computed occupation numbers, we initiate Q2FMM from the finest level, where each box represents one lattice site. At this level, the time evolution is computed by the interaction between individual qubits that are in each other's interaction list. For the coarser levels, the occupation number of the boxes is obtained by summing the occupation numbers of their children. To avoid the premature erasure of information stored in ancilla qubits from previous levels, we defer uncomputing until all evolving phases have been evaluated. 
Therefore, the full quantum circuit can be partitioned into a compute phase and an uncompute phase. In the compute phase, we start from the finest level, where interactions are restricted to single grid points, and we apply the time evolution operator directly. 
For each of the other levels, we proceed as follows:
\begin{enumerate}
    \item We first obtain the occupation number of each box at the current level by summing the values of its four children using quantum adders.
    \item Subsequently, we implement the evolution in Eq.~\eqref{eq:time_l} with these occupation numbers by means of the quantum multiplier.
\end{enumerate}
This procedure is recursively repeated to traverse the entire hierarchy until the coarsest level is reached, as shown in Fig.\ref{fig:point1}. Note that after the summation, every four child boxes are efficiently merged into their corresponding parent box. Since the quantum adders and multipliers are reversible, the uncomputing merely reverses this merging procedure while leaving the accumulated phase intact, so it is safe to perform. As illustrated in Fig.~\ref{fig:box_uncompute}, uncomputing is carried out in the reverse order of the compute phase: at each step, the occupation number stored for a parent box $P$ is decomposed back into those of its four child boxes $C$ at the next finer level. As a concrete example, a quantum circuit for the 0th-order Q2FMM is presented in Appendix~\ref{app:overall_circuit}, shown for a one-dimensional lattice for visual clarity.

\begin{figure}[t!]
  \centering
  \resizebox{1\columnwidth}{!}{%
\begin{tikzpicture}[>=stealth, line cap=round, line join=round]

  \definecolor{deepgreen}{RGB}{0,100,0}

  \draw[step=2cm, line width=1.6pt, color=deepgreen] (0,-8.001) grid (6,-1.999);

  \node[font=\LARGE] at (1,-7) {P};

  \draw[
    step=1cm,
    line width=1.0pt,
    color=black,
    dash pattern=on 2.5pt off 2pt
  ] (8,-8) grid (14,-2);

  \draw[line width=1.6pt, color=deepgreen] (8,-8) rectangle (10,-6);
  \draw[line width=1.6pt, color=deepgreen] (10,-8) rectangle (12,-6);
  \draw[line width=1.6pt, color=deepgreen] (12,-8) rectangle (14,-6);

  \draw[line width=1.6pt, color=deepgreen] (8,-6) rectangle (10,-4);
  \draw[line width=1.6pt, color=deepgreen] (10,-6) rectangle (12,-4);
  \draw[line width=1.6pt, color=deepgreen] (12,-6) rectangle (14,-4);

  \draw[line width=1.6pt, color=deepgreen] (8,-4) rectangle (10,-2);
  \draw[line width=1.6pt, color=deepgreen] (10,-4) rectangle (12,-2);
  \draw[line width=1.6pt, color=deepgreen] (12,-4) rectangle (14,-2);

  \node[font=\Large] at (8.5,-7.5) {C};
  \node[font=\Large] at (9.5,-7.5) {C};
  \node[font=\Large] at (8.5,-6.5) {C};
  \node[font=\Large] at (9.5,-6.5) {C};

  \draw[->, line width=1.2pt] (6.2,-5) -- (7.8,-5)
      node[midway, above, align=center, font=\Large] {un-\\compute}
      node[midway, below, font=\Large] {finer};

  \draw[->, line width=1.5pt] (14.2,-5) -- (15.8,-5)
      node[midway, above, align=center, font=\Large] {un-\\compute}
      node[midway, below, font=\Large] {finer};

\end{tikzpicture}%
}
  \caption{Uncomputing proceeds from the coarsest level to the finest level, i.e., in the reverse order of the compute phase. Here, $P$ denotes a parent box at a given level $L$, and the four boxes labeled $C$ denote its child boxes at the next finer level $L+1$.}
  \label{fig:box_uncompute}
\end{figure}

The above construction addresses the main concerns raised in implementing the FMM for time evolution on a quantum computer. It shows how hierarchical information can be reused and how to get rid of the effects from garbage states or unwanted entanglement through uncomputing. The $0^\text{th}$-order Q2FMM already captures the main principles of the full Q2FMM. In the higher-order setting, the overall hierarchical workflow and circuit structure remain essentially the same. The main modification is that the subroutines for aggregating the information of smaller boxes into larger ones and for evaluating the interactions between boxes must be replaced by their higher-order counterparts, which retain the more detailed internal structure of each region. In this sense, the higher-order construction can be viewed as a refinement of the same framework, rather than a fundamentally different algorithm.

\subsection{Circuit depth for $0^\text{th}$-order Q2FMM}
\label{sec:0th_cost}

Before turning to the higher-order Q2FMM, we first briefly analyze the circuit depth of the $0^\text{th}$-order scheme. This not only highlights why our algorithm achieves favorable complexity but also provides intuition for the higher-order case, whose scaling follows a similar structure. 

The main contributions to the circuit depth come from three components: (i) the quantum adders used to compute the occupation numbers, (ii) the quantum multipliers used to compute the products of occupation numbers, and (iii) the phase gates used to implement the resulting phases. Among these, the second component clearly dominates and therefore determines the overall depth scaling, since quantum multipliers are more costly than quantum adders, while the phase gates can be implemented in parallel in constant depth.

Therefore, the main task is to estimate the depth of quantum multipliers. We assume an input length of $n = \log_2(Q+1)$, corresponding to the largest possible value in our calculation, where $Q$ denotes the number of electrons of the system. 
Selecting an efficient quantum multiplier is crucial for maintaining a favorable circuit depth scaling. 
For instance, the depth of the quantum multiplier proposed by Kahanamoku-Meyer et al.~\cite{kahanamoku2024fast} is bounded by $\mathcal{O}(D_{\text{QFT}}(n))$ if a few ancilla qubits are allowed, where $D_{\text{QFT}}$ is the circuit depth of the quantum Fourier transform (QFT) for $n$-bit input; selecting such a quantum multiplier leads to depth $\mathcal{O}(\log Q)$ if we consider the standard QFT circuit implementation~\cite{maslov2007lineardepth}, and other advanced implementations could further reduce the depth~\cite{kahanamoku-meyer2025optimisticqft, cleve2000fast, barenco1996aqft}. Since there are $\mathcal{O}(\log N)$ levels, the overall circuit depth of the algorithm exhibits a polylogarithmic scaling, $\mathcal{O}(\log N \cdot \log Q)$, when the quantum multiplier of Ref.~\cite{kahanamoku2024fast} is employed.

\subsection{Higher-order FMM}
\label{sec:higher_FMM}

\begin{figure}[b]
\centering
\begin{tikzpicture}

  \draw (0,0) -- (3.6,-1.6);

  \draw[->, >=stealth', red!50!black, line width=1pt] (0,0) -- (-0.6,0.8) ;
  \draw[->, >=stealth', red!50!black, line width=1pt] (3.6,-1.6) -- (4.2,-0.8);

  \node at (0.2,0.2) {$A$};
  \node at (3.5,-1.2) {$B$};

  \fill[red!50!black] (0,0) circle (1.5pt);
  \fill[red!50!black] (3.6,-1.6) circle (1.5pt);

  \node at (-0.8,0.8) {$a$};
  \node at (4.4,-0.8) {$b$};
  
  \draw [dashed](-0.6,0.8) -- (4.2,-0.8);

  \node at (-0.55,0.3) {$\mathbf{r}_{aA}$};
  \node at (4.3,-1.3) {$\mathbf{r}_{bB}$};
  \node at (1.5,-1.0) {$\mathbf{r}_{AB}$};
  \node at (2.0,0.2) {$\mathbf{r}_{ab}$};

\end{tikzpicture}
\caption{The expression that originally depends on the two target points $a$ and $b$ is rewritten in terms of local contributions around the centers $A$ and $B$, together with the interaction between the two centers.}
\label{fig:expansion}
\end{figure} 

In this subsection, we extend Q2FMM to higher-order FMM, which is essential for improving the accuracy systematically. Conceptually, this extension does not change the overall algorithmic workflow: the same hierarchical compute-reuse-uncompute structure is retained, while the information stored for each box is refined from a single occupation number to a collection of multipole coefficients. In this sense, the higher-order Q2FMM should be viewed as a refinement of the same framework, rather than as a fundamentally different algorithm. We now introduce the corresponding higher-order FMM formulation.

The FMM formulation for the Coulomb kernel $K_{\text{C}}(\mathbf{r}_a,\mathbf{r}_b) \coloneqq 1/\|\mathbf{r}_a - \mathbf{r}_b\|_2$ is based on its multipole expansion~\cite{helgaker2000molecular}. We employ the regular and irregular solid harmonics~\cite{stone2013intermolecular}, defined as
\begin{subequations}
\begin{align}
\label{eq:r_lm}
R_{\ell m}(\mathbf{r}) &\coloneqq 
    \sqrt{(\ell-m)!\,(\ell+m)!}\; r^{\ell}\, C_{\ell m}(\theta,\phi),
\end{align}
\begin{align}
I_{\ell m}(\mathbf{r}) &\coloneqq 
    \sqrt{(\ell-m)!\,(\ell+m)!}\; r^{-(\ell+1)}\, C_{\ell m}(\theta,\phi),
\end{align}
\end{subequations}
where $C_{\ell m}(\theta,\phi)$ are the Racah-normalized spherical harmonics. Using these definitions, the Coulomb kernel can be expanded by the solid harmonic addition theorem~\cite{brink1968angular} as
\begin{equation}
\label{eq:kernel_expansion}
\begin{split}
K_{\text{C}}(\mathbf{r}_a,\mathbf{r}_b)
&= \sum_{\ell=0}^{\infty} \sum_{m=-\ell}^{\ell} 
   \sum_{j=0}^{\infty} \sum_{k=-j}^{j} 
   (-1)^{j} \,
   R_{\ell m}(\mathbf{r}_{aA}) \\
&\quad \times
   I^{*}_{\ell+j,\, m+k}(\mathbf{r}_{AB}) \,
   R_{jk}(\mathbf{r}_{bB}),
\end{split}
\end{equation}
where $I^{*}(\mathbf{r}_{AB})$ denotes the complex conjugate of $I(\mathbf{r}_{AB})$, and $\mathbf{r}_{AB} = \mathbf{r}_A - \mathbf{r}_B$ denotes the displacement between two centers. We refer to the textbook by Helgaker et al.~\cite{helgaker2000molecular} for a detailed derivation. This formula expresses $K_{\text{C}}(\mathbf{r}_a,\mathbf{r}_b)$ in terms of local coordinates $\mathbf{r}_{aA}=\mathbf{r}_a-\mathbf{r}_A$ and 
$\mathbf{r}_{bB}=\mathbf{r}_b-\mathbf{r}_B$ relative to
expansion centers $A$ and $B$, as shown in Fig.~\ref{fig:expansion}. Note that $\ell = j = 0$ is exactly the zeroth-order case we discussed.

With Eq.~\eqref{eq:kernel_expansion}, and assuming two well-separated boxes $A$ and $B$, the box-box interaction in the higher-order FMM can be calculated by summing over all the interaction pairs:
\begin{equation}
\begin{aligned}
\label{eq:expansion_box}
E_{AB}= &\sum_{a \in A} \sum_{b \in B} K_{\text{C}}(\mathbf{r}_a,\mathbf{r}_b) {q_a q_b}\\
\approx & \sum_{\ell=0}^{p}  \sum_{m=-\ell}^{\ell} \sum_{j=0}^{p-\ell} \sum_{k=-j}^{j} \left(\sum_{a \in A} R_{\ell m}(\mathbf{r}_{aA}) q_a\right) \\
\times & I^{*}_{\ell+j,\, m+k} (\mathbf{r}_{AB}) \, \left(\sum_{b \in B}R_{jk}(\mathbf{r}_{bB}) q_b\right)(-1)^{j}.
\end{aligned}
\end{equation}
Instead of representing each box only by its total occupation number, we now retain controlled information about its internal structure through multipole coefficients.
The infinite series over $\ell$ and $j$ in Eq.~\eqref{eq:kernel_expansion} is truncated such that $\ell+j \le p$, where $p$ denotes the FMM truncation order, controlling the trade-off between accuracy and computational cost. Here, $q_a \in \{0,1\}$ represents the electron number at point $a$. Assuming that each box contains $\kappa$ grid points, Eq.~\eqref{eq:expansion_box} reduces the computational effort from evaluating all $\kappa^2$ point-to-point interactions to only $\mathcal{O}(\kappa (p+1)^2)$ local terms, as shown in Fig.~\ref{fig:expansion1}. Note that the case $p=0$ corresponds exactly to the $0^\text{th}$-order FMM discussed in Sec.~\ref{sec:0th_q2fmm}. The expansion converges for well-separated boxes, i.e.,
\begin{equation}
\max\{\norm{\mathbf{r}_{aA}}_2, \norm{\mathbf{r}_{bB}}_2\} < \norm{\mathbf{r}_{AB}}_2,
\end{equation}
and the interaction-list criterion in our hierarchical structure automatically satisfies this condition. 
Incorporating Eq.~\eqref{eq:expansion_box} into the hierarchical framework discussed in Sec.~\ref {sec:0thcircuit} yields the higher-order Q2FMM, in which box-box interactions are evaluated using Eq.~\eqref{eq:expansion_box}. 

\subsection{Implementation of higher-order Q2FMM on a quantum computer}

Therefore, the overall procedure is analogous to the $0^\text{th}$-order case: we first evaluate the energy on the second-finest level, where each box contains only four grid points. From there, the multipole information is translated and aggregated to coarser levels. At each level, we compute box-box interactions, perform the time-evolution, and then proceed to the next coarser level, traversing the hierarchy from finest to coarsest. The main difference is that the box information is now given by multipole coefficients rather than by a single occupation number.
Replacing $q_a \mapsto \hat{n}_a$ in Eq.~\eqref{eq:expansion_box} defines the interaction operator $V_{AB}$, which consists solely of products of number operators, as in the $0^\text{th}$-order case. Consequently, any configuration $\ket{c_{AB}}$ (i.e., computational basis state of box $A$ and $B$) is an eigenstate of $V_{AB}$:
\begin{equation}
    V_{AB} \ket{c_{AB}} = E^{c}_{AB}\ket{c_{AB}},
\end{equation}
where $E^{c}_{AB}$ denotes the corresponding eigenvalue computed via Eq.~\eqref{eq:expansion_box}. 
The time evolution governed by $V_{AB}$ therefore results only in phase changes for each configuration $\ket{c_{AB}}$:
\begin{equation}
    e^{-itV_{AB}}\ket{c_{AB}} = e^{-itE^{c}_{AB}}\ket{c_{AB}}.
\end{equation}
Hence, our task again reduces to evaluating the Coulomb potential between $A$ and $B$ for each configuration using Eq.~\eqref{eq:expansion_box} and generating the corresponding evolving phase through single-qubit projector gates, as discussed in Eq.~\eqref{eq:phase} for the $0^\text{th}$-order case.

\begin{figure}
\centering
\begin{tikzpicture}

  \draw (0,0) -- (3.6,-1.6);

  \draw (0,0) -- (-0.4,0.8);
  \fill[red!50!black] (-0.4,0.8) circle (2pt); 

  \draw (0,0) -- (-0.7,-0.3);
  \fill[red!50!black] (-0.7,-0.3) circle (2pt); 

  \draw (0,0) -- (-0.7,-0.7);
  \fill[red!50!black] (-0.7,-0.7) circle (2pt); 

  \draw (0,0) circle [radius=1.1cm];

  \draw (3.6,-1.6) -- (4.2,-0.8);
  \fill[red!50!black] (4.2,-0.8) circle (2pt); 

  \draw (3.6,-1.6) -- (4.0,-1.5);
  \fill[red!50!black] (4.0,-1.5) circle (2pt); 

  \draw (3.6,-1.6) -- (3.7,-2.5);
  \fill[red!50!black] (3.7,-2.5) circle (2pt); 

  \draw (3.6,-1.6) circle [radius=1.1cm];

  \node at (0.2,0.2) {$A$};
  \node at (3.5,-1.3) {$B$};

  \fill[red!50!black] (0,0) circle (1.5pt);
  \fill[red!50!black] (3.6,-1.6) circle (1.5pt);

\end{tikzpicture}
\caption{Illustration of higher-order FMM, Eq.~\eqref{eq:expansion_box}: The overall pairwise sum is obtained by first calculating per-box local information and then evaluating box-box interactions.}
\label{fig:expansion1}
\end{figure}

As discussed for the $0^\text{th}$-order FMM in Sec.~\ref{sec:0th_q2fmm}, box-box interactions at each level are obtained from calculations performed on the immediately finer level. To illustrate this procedure, we begin with the second finest level, where each box contains four grid points of the physical Fermi-Hubbard lattice, and consequently discuss how the computed information propagates to coarser levels. We first simplify Eq.~\eqref{eq:expansion_box} by defining the multipole expansion $M^A_{\ell m}$ and $M^B_{jk}$:
\begin{subequations}
\label{eq:M}
\begin{align}
\label{eq:MA}
    M^A_{\ell m} \coloneqq \sum_{a \in A} \,
   R_{\ell m}(\mathbf{r}_{aA}) q_a,
\end{align}
\begin{align}
\label{eq:MB}
    M^B_{jk} \coloneqq \sum_{b \in B} \,
   R_{j k}(\mathbf{r}_{bB}) q_b,
\end{align}
\end{subequations}
Here, the indices $(\ell,m)$ and $(j,k)$ follow the truncation in Eq.~\eqref{eq:kernel_expansion}. Thus, Eq.~\eqref{eq:expansion_box} can be rewritten as:
\begin{equation}
\label{eq:Eab}
E_{AB} \approx \sum_{\ell=0}^{p} \sum_{m=-\ell}^{\ell} \sum_{j=0}^{p-\ell} \sum_{k=-j}^{j} \,{M^A_{\ell m}} 
I^{*}_{\ell +j,\, m+k}(\mathbf{r}_{AB}) \, {M^B_{jk}} (-1)^{j}.
\end{equation}

An efficient way to compute $M^A_{\ell m}$ and $M^B_{jk}$ in Eq.~\eqref{eq:M} is to encode the binary representation of $R_{\ell m} (\mathbf{r}_{aA})$, classically precomputed by Eq.~\eqref{eq:r_lm}, as a sequence of CNOT gates conditioned on $\ket{q_a}$. As illustrated in Fig.~\ref{fig:rq_circuit}, if the register storing the result of $R_{\ell m} (\mathbf{r}_{aA}) q_a$ is initialized to $\ket{00\cdots0}$, applying these CNOT gates yields the final state that correctly encodes the product $R_{\ell m}(\mathbf{r}_{aA})q_a$. Because computational basis states natively encode integers, we use fixed-point scaling for non-integer values; for example, for a decimal $b_1 b_2.b_3 b_4$, we store $b_1 b_2 b_3 b_4$ and carry a scale factor $10^{-2}$; equivalently, the extra factor $10^{-2}$ is absorbed into the time duration during the time evolution step. The quantity $M^A_{\ell m}$ is then obtained by accumulating the $R_{\ell m}(\mathbf{r}_{aA})q_a$ terms using quantum adders. 
After obtaining $M^A_{\ell m}$ and $M^B_{jk}$, the evaluation of $E_{AB}$ in Eq.~\eqref{eq:Eab} becomes straightforward: each term in the sum is computed with quantum multiplication and accumulated with quantum addition. Similar to the $0^\text{th}$-order Q2FMM, the resulting time evolution phase is obtained by applying the phase gates as defined in Eq.~\eqref{eq:sequence_p}.
Note that, for a fixed FMM order $p$, the depth of this stage is independent of the system size $N$.

This second-finest level serves as the base case of the higher-order hierarchical procedure. Once the multipole information for these smallest nontrivial boxes has been prepared, the corresponding information for coarser levels is obtained by translation and aggregation, which we discuss next.

\begin{figure}[t] 
  \centering
  \begin{quantikz}[row sep=0.1cm]
\lstick{$\ket{q_a}$} & \ctrl{1} & \ctrl{2}    & \ctrl{4} & \ghost{\targ{}}\\
\lstick{$\ket{0}$} & \targ{}  & \qw              & \qw      & \ghost{\targ{}}\\
\lstick{$\ket{0}$} & \qw      & \targ{}         & \qw      & \ghost{\targ{}}\\
\lstick{$\ket{0}$} & \qw      & \qw         & \qw      & \ghost{\targ{}}\\
\lstick{$\ket{0}$} & \qw      & \qw              & \targ{}  & \ghost{\targ{}}
\end{quantikz}
  \caption{Computation of $R_{\ell m}(\mathbf{r}_{aA}) q_a$ using CNOT gates, illustrated for the case where $R_{\ell m}(\mathbf{r}_{aA})$ is encoded in the integer form $\ket{1101}$. A CNOT gate is applied to each qubit corresponding to a nonzero bit of $R_{\ell m}$. When $q_a=0$, the output state is $\ket{0000}$; when $q_a=1$, the output state is $\ket{1101}$, both of these two cases correctly encoding the product $R_{\ell m} q_a$. Note that we might need more qubits to reach the desired precision.}
  \label{fig:rq_circuit}
\end{figure}

\subsection{Reusing multipole expansions across levels}
\label{sec:m_to_m}

Similar to the $0^\text{th}$-order Q2FMM discussed in Sec.~\ref{sec:0thcircuit}, computations for coarser levels can leverage information from the child boxes. 
For the higher-order case, the parent box's multipole expansion is obtained by applying a multipole-to-multipole (M2M) translation to each child. In other words, the child box's multipole expansion $M_{\ell m}$ can be reused by shifting its expansion origin from the child's center to the parent's center. The parent box's multipole expansion is then obtained by summing the shifted expansions from all child boxes. 
For example, let $\mathbf{x}_C$ and $\mathbf{x}_P$ be the centers of a child $C$ and its parent $P$, respectively, and define the displacement
$\mathbf{d} \coloneqq \mathbf{x}_P - \mathbf{x}_C$ (from child to parent).
The translation from the multipole expansion of the child box $M^C_{\ell m}$ to the one of parent center $M^{C\rightarrow P}_{\ell m}$ can be obtained by the addition theorem~\cite{helgaker2000molecular}:
\begin{equation}
\label{eq:translation}
    M^{C\rightarrow P}_{\ell m} = \sum_{j=0}^\ell \sum_{k=-j}^j R_{\ell -j,m-k}(-\mathbf{d}) M^C_{jk}.
\end{equation}
A parent's total multipole expansions are then obtained by summation over all its child boxes' contributions:
\begin{equation}
\label{eq:c_to_p}
    M^{P}_{\ell m} = \sum_{C} M^{C\rightarrow P}_{\ell m}.
\end{equation}
This process is repeated each time we reach a coarser level. The Coulomb interaction between parent boxes is then computed from their multipole expansions as given in Eq.~\eqref{eq:Eab}. Similar to the $0^\text{th}$-order Q2FMM, uncomputing is needed to transfer the accumulated phases back to the qubits. In practice, we defer the uncomputing to the end of the procedure so that the computed multipole expansions can be reused at the next coarser level, just as in the $0^\text{th}$-order scheme. In summary, at the algorithmic level, the higher-order Q2FMM preserves the same compute-reuse-uncompute pattern as the $0^\text{th}$-order construction; only the data carried by each box and the corresponding local update rules are refined.

\subsection{Error analysis}
\label{sec:error}

The Q2FMM algorithm involves three sources of error: the FMM error $\epsilon_F$, the rounding error $\epsilon_b$, and the Trotter error $\epsilon_T$. Since the resources required to implement a quantum algorithm depend directly on the target accuracy, error analysis is essential.

We first estimate the error introduced by the FMM approximation, which also allows us to determine the required FMM order $p$ for achieving a desired error. For an arbitrary state $\ket{\psi} = \sum_i \alpha_i \ket{i}$, the higher-order Q2FMM introduces an error $\epsilon^{i}_F$ to the overall Coulomb interaction for a given Fock basis state $\ket{i}$. When computing the time evolution governed by the Coulomb interaction, such an energy error leads to the time evolution error of basis state $\ket{i}$:
\begin{equation}
\begin{aligned}
    \abs*{e^{-i\delta t(E^i_C+\epsilon^{i}_F)}-e^{-i\delta t E^i_C}}
    &= \abs*{e^{-i \delta tE^i_C}} \abs*{e^{-i \delta t\epsilon^{i}_F}-1} \\
    &= \abs*{e^{-i\delta t\epsilon^{i}_F}-1} \\
    &\le \delta t \, \abs{\epsilon^{i}_F}
\end{aligned}
\end{equation}
with a small $\delta t$, where we denote the overall Coulomb interaction of $\ket{i}$ by $E^i_C$. Here, the quantities $e^{-i\delta t(E^i_C+\epsilon^{i}_F)}$ and $e^{-i\delta t E^i_C}$ are complex numbers, and $|\cdot|$ was used to denote the complex modulus of the difference.
Therefore, the overall time evolution error incurred by Q2FMM for an arbitrary state $\ket{\psi} = \sum_i \alpha_i \ket{i}$ is bounded by:
\begin{equation}
    \epsilon_{F,\delta t} \le \sum_i \abs{\alpha_i}^2 \, \abs{\epsilon^{i}_F} \, \delta t 
\end{equation}
The FMM error $\epsilon^i_F$ scales as $\mathcal{O}\left(\frac{r}{R}^{p+1}\right)$~\cite{greengard1987fast, helgaker2000molecular}, where $R$ denotes the center-center distance between interacting boxes and $r$ the box radius at a given level. Hence, the time evolution error scales as: 
\begin{equation}
    \epsilon_{F,t} \sim \mathcal{O}\left(Q\, t \,  \left( \frac{r}{R} \right)^{p+1}\right)
\end{equation}
for a total time duration $t$. Importantly, this error decreases rapidly with increasing expansion order $p$, exhibiting geometric convergence. From the above equation, we can also see that the truncation order $p$ scales as:
\begin{equation}
    p \sim \mathcal{O}(\log(Q\, t \,/\epsilon_{F,t})).
\end{equation} 
which means $p$ scales logarithmically with total electron number $Q$, the target error $\epsilon_{F,t}$ and evolution time $t$.

We also note that the dependence on the total charge $Q$ could be treated more mildly if one assumes that the electron distribution satisfies the standard bounded-density condition, i.e., the distribution is reasonably uniform. In this case, the FMM error can be replaced by $\epsilon_{F,t} \sim \mathcal{O}\left(g \, t \,  \left( \frac{r}{R} \right)^{p+1}\right)$, where $g$ is a constant of order $\mathcal{O}(1)$~\cite{berry2025qfmm}. However, this will not affect the final asymptotic scaling of this work.

Then, we estimate the Trotter error $\epsilon_T$, which is essential for determining the number of Trotter steps needed. According to the commutator-based theory of Trotter error~\cite{Childs2021PRXCommutator}, the total Trotter error scales as:
\begin{equation}
\label{eq:trotter_error}
    \epsilon_T = \mathcal{O}(t \, \delta t^2 \Lambda),
\end{equation}
where:
\begin{equation}
\label{eq:commutator}
    \Lambda \le \mathcal{O}( N \, \log^2 N)
\end{equation}
for the extended Hubbard model considered here; see Appendix~\ref{app:commutator} for derivation details. Here, we omit the Hamiltonian coefficients, as our main interest lies in the asymptotic scaling with the system size $N$.

Finally, since Q2FMM employs quantum arithmetic circuits, it naturally involves an additional source of error due to finite-precision arithmetic, namely, the rounding error $\epsilon_b$.
It is straightforward to see that
\begin{equation}
    \epsilon_b = \mathcal{O}(2^{-n_d}),
\end{equation}
where $n_d$ denotes the number of bits after the decimal point. Hence, to reach the binary accuracy $\epsilon_b$,
\begin{equation}
    n_d = \log_2(1/\epsilon_b)
\end{equation}
bits are needed after the decimal point.

\subsection{Circuit depth of higher-order Q2FMM}
\label{sec:resources}

Following the outline presented in Sec.~\ref{sec:0th_cost} for $0^\text{th}$-order Q2FMM, the circuit depth for higher-order Q2FMM is similarly dominated by quantum arithmetic circuits, especially the quantum multipliers. Concretely, Eqs.~\eqref{eq:Eab}-~\eqref{eq:c_to_p} are implemented with quantum adders and quantum multipliers. Note that Eq.~\eqref{eq:translation} is local to each child box and can be executed in parallel across children.
For each child, and with fixed truncation order $p$, executing both Eq.~\eqref{eq:Eab} and Eq.~\eqref{eq:translation} requires $\mathcal{O}(p^4)$ multiplications. In other words, the number of multiplications scales logarithmically.

The depth of the quantum multiplier used in Eq.~\eqref{eq:Eab} and Eq.~\eqref{eq:translation} is governed by the input length, i.e., the bit length of the ancilla register that stores the multipole coefficients $M^A_{\ell m}$ for box $A$, which satisfy the bound:
\begin{equation}
\begin{split}
\big|M^A_{\ell m}\big| &= \Big|\sum_{i=1}^{Q_b} q_i\, R_{\ell m}(\mathbf r_{iA})\Big| \\
&\le\; \sum_{i=1}^{Q_b} |q_i|\, \big|R_{\ell m}(\mathbf r_{iA})\big| \\
&\le\; Q_b \max_{1\le i\le Q_b}\big|R_{\ell m}(\mathbf r_{iA})\big| \\
&\le\; Q_b \sqrt{(\ell-m)!\,(\ell+m)!}\; r_A^{\,\ell},
\end{split}
\end{equation}
where $r_A$ is the radius of box $A$ (the upper bound of $r_{iA}$), and $Q_b$ denotes the number of electrons in a given box. Here, $|\cdot|$ also denotes the complex modulus. After normalizing $M^{A}_{\ell m}$ by
$\sqrt{(2\ell)!}\, r_A^\ell$, the resulting quantity is bounded by $Q$ and can be encoded as a binary fixed-point number. Hence, to represent $M^A_{\ell m}$ within binary precision $\epsilon_b$, we need at most:
\begin{equation}
    n_{\epsilon_b, \text{max}} =  \log_2(Q+1)+\log_2(1/\epsilon_b)
\end{equation}
bits, where the first term encodes the integer range and the second term encodes the binary fractional precision. 

Using the efficient multiplier proposed by Kahanamoku-Meyer et al.~\cite{kahanamoku2024fast}, the circuit depth \emph{per level} scales asymptotically as $\mathcal{O}(p^4(\log Q+\log({1}/{\epsilon_b})))$. Since $p$ scales logarithmically and there are $\log (N)$ levels in total, the total depth for simulating a single Trotter step remains \emph{polylogarithmic}. For instance, when the quantum multiplier of Ref.~\cite{kahanamoku2024fast} is selected, the overall circuit depth scales as $\mathcal{O}(\log(N) \cdot (\log Q+\log({1}/{\epsilon_b})))$; other widely known quantum multipliers, such as QFT-based multipliers~\cite{ruizperez2017qftarith} also help reach logarithmic depth scaling. Here, we only consider per-step depth; we discuss overall depth later.

\subsection{Gate complexity of higher-order Q2FMM}
The gate complexity of the higher-order Q2FMM can be estimated by summing the gates required at each hierarchical level. As mentioned earlier, the quantum multiplier is the dominant cost contributor. Thus, focusing solely on the cost of the quantum multipliers is sufficient to determine the asymptotic scaling of gate complexity. Let the number of bits required to store a single coefficient with absolute precision $\epsilon_b$ be
\begin{equation}
    n_{\epsilon_b, \beta} = \log_2(\beta+1) + \log_2(1/\epsilon_b),
\end{equation}
where $\beta \in \{1, \dots, 4^{L_{\text{max}}-2}\}$ denotes the number of grid points contained in a box at a given level (so that $\beta=1$ corresponds to the finest level). Note that $n_{\epsilon_b, \beta}$ is capped at $\log_2(Q+1) + \log_2(1/\epsilon_b)$. Still selecting the quantum multiplier in Ref.~\cite{kahanamoku2024fast}, its gate complexity scales as $\mathcal{O}(n^{1.3})$ with input length $n$. Considering there are $N/\beta$ boxes at a level with box size $\beta$, the total number of gates scales asymptotically as:
\begin{equation}
\begin{aligned}
    N_G &= \sum_\beta \frac{p^4 N}{\beta} \cdot \mathcal{O}(n_{\epsilon_b,\beta})^{1.3}  \\
    &= \mathcal{O}(N \,p^4 \cdot \log^{1.3}(1/\epsilon_b))\\
    &= \mathcal{O}(N \,\log^4(N \,t \,/\epsilon_{F,t}) \cdot \log^{1.3}(1/\epsilon_b))
\end{aligned}
\end{equation}
and we can further simplify the number of gates as:
\begin{equation}
    N_G = \Tilde{\mathcal{O}}(N),
\end{equation}
which implies that the per-step gate complexity scales linearly with system size $N$ with given $\epsilon_b$ and $\epsilon_{F,t}$. Here, the notation $\Tilde{\mathcal{O}}(\cdot)$ suppresses polylogarithmic factors.

Having estimated the circuit depth and gate complexity per Trotter step, we now turn to the overall complexity for simulating a total evolution time $t$ with target Trotter error $\epsilon_T$, which requires accounting for the total number of Trotter steps. First, following Eq.~\eqref{eq:trotter_error}, the Trotter step size scales as:
\begin{equation}
    \delta t = \mathcal{O}\left(\frac{\epsilon^{1/2}_T}{t^{1/2} \, \Lambda^{1/2}}\right).
\end{equation}
Accordingly, the number of Trotter steps scales as:
\begin{equation}
    \Omega = \frac{t}{\delta t} = \mathcal{O}\left(\frac{t^{3/2} \Lambda^{1/2}}{\epsilon^{1/2}_T} \right).
\end{equation}
Substituting Eq.~\eqref{eq:commutator} gives the total number of Trotter steps:
\begin{equation}
    \Omega = \mathcal{O}\left(
    \frac{t^{3/2} N^{1/2}\,\log N}{\epsilon^{1/2}_T}
    \right) = \Tilde{\mathcal{O}}\left(
    \frac{t^{3/2} N^{1/2}}{\epsilon^{1/2}_T}
    \right).
\end{equation}
Therefore, the total circuit depth also scales as:
\begin{equation}
    \Tilde{\mathcal{O}}\left(
    \frac{t^{3/2} N^{1/2}}{\epsilon^{1/2}_T}
    \right)
\end{equation}
since the circuit depth for a single Trotter step scales polylogarithmically. Likewise, the total gate complexity scales as:
\begin{equation}
    \Tilde{\mathcal{O}}\left(\frac{t^{3/2} N^{3/2}}{\epsilon^{1/2}_T}\right)
\end{equation}
because the gate complexity of a single Trotter step scales as $\Tilde{\mathcal{O}}(N)$. The dependencies of both the FMM error $\epsilon_F$ and the binary approximation error $\epsilon_b$ are polylogarithmic, and are therefore suppressed in the $\Tilde{\mathcal{O}}(\cdot)$ notation.

The above resource estimate is based on a common circuit model and counts only logical primitive gates, as is standard in gate-complexity analyses for Hamiltonian simulation. Nevertheless, we also would like to emphasize that the long-range operations considered here, i.e., the gates applied to two distant boxes, are not merely a fictitious idealization. For instance, it has been shown that long-range gates can be implemented in constant depth using the surface code, with only nearest-neighbor connectivity~\cite{litinski2019game, edp2022Beverland}. Moreover, long-range operations can also be realized by shuttling the target quantum register to the same region and then applying only local gates~\cite{bluvstein2022processor,bluvstein2024logical,evered2023high,schmid2024compiler}, as illustrated in Appendix~\ref{app:shuttle}.

\subsection{Ancilla qubits requirement of higher-order Q2FMM}
Finally, we estimate the number of ancilla qubits required for Q2FMM, which is dominated by those used to store the multipole expansions. Given $M_{\ell m}$ contains $(p+1)^2$ coefficients, the total number of ancilla qubits needed at a given level scales as
\begin{equation}
\begin{aligned}
N_s &\propto \frac{N}{\beta}(p+1)^2\,n_{\epsilon_b,\beta} \\
    &= \frac{N}{\beta}(p+1)^2\bigl(\log_2(\beta+1)+\log_2(1/\epsilon_b)\bigr),
\end{aligned}
\end{equation}
recall that  $\beta \in \{1, \dots, 4^{L_{\text{max}}-2}\}$ denotes the number of grid points contained in a box at this level.
Even under a large error tolerance $\epsilon_b \le 10^{-3}$ to represent a decimal, a direct maximization shows that $N_s$ is largest at the finest level $\beta=1$. In this case,
$n_{\epsilon_b, 1}=1+\log_2(1/\epsilon_b)$, so storing all $(p+1)^2$ terms in $M_{\ell m}$ for one site requires $(p+1)^2\bigl(1+\log_2(1/\epsilon_b)\bigr)$ qubits, and the overall ancilla qubits requirement is:
\begin{equation}
\label{eq:n_anci}
    N_{\text{ancilla}} = N\,(p+1)^2\bigl(1+\log_2(1/\epsilon_b)\bigr) = \Tilde{\mathcal{O}}(N),
\end{equation}
i.e., the total number of ancilla qubits scales quasi-linearly with the number of sites $N$ for the target accuracy. 

\subsection{Comparison with the method based on the interaction picture}

\begin{table*}
    \centering
    \caption{Comparison between Q2FMM and the method based on the interaction picture~\cite{low2018interactionpicture}. Here, we assume that the different error contributions are balanced and each is bounded by $\epsilon$.}
    \label{tab:q2fmm_low_wiebe_comparison}
    \begin{ruledtabular}
    \begin{tabular}{@{}l@{\hspace{2pt}}c@{\hspace{2pt}}c@{\hspace{2pt}}c@{\hspace{2pt}}c@{\hspace{2pt}}c@{}}
        Method & Depth & Gate & Ancillae & Efficient w/ PBC & Efficient w/ OBC \\
        Q2FMM 
        & $\Tilde{\mathcal{O}}(t^{\frac{3}{2}}N^{\frac{1}{2}}/\epsilon^{\frac{1}{2}})$ 
        & $\Tilde{\mathcal{O}}(t^{\frac{3}{2}}N^{\frac{3}{2}}/\epsilon^{\frac{1}{2}})$ 
        & $\Tilde{\mathcal{O}}(N)$ 
        & w/ periodic FMM 
        & Yes \\
        Ref.~\cite{low2018interactionpicture} 
        & $\Tilde{\mathcal{O}}(tN)$ 
        & $\Tilde{\mathcal{O}}(tN^2)$ 
        & $\Tilde{\mathcal{O}}(N)$ 
        & Yes 
        & w/ zero-padding \\
    \end{tabular}
    \end{ruledtabular}
\end{table*}

We note that the interaction-picture algorithm in the block-encoding framework proposed by G.~H.~Low and N.~Wiebe~\cite{low2018interactionpicture} achieves a low overall gate complexity of $\Tilde{\mathcal{O}}(N^2 t)$ for the extended Hubbard model. By comparison, Q2FMM offers two potential advantages or differences, as discussed below and summarized in Table~\ref {tab:q2fmm_low_wiebe_comparison}.

First, our approach yields an overall gate complexity of $\Tilde{\mathcal{O}}(t^{3/2} N^{3/2}/\epsilon^{1/2})$, where $\epsilon$ denotes the target accuracy, under the assumption that the different error contributions are balanced and each is bounded by $\epsilon$. Compared with the gate complexity $\Tilde{\mathcal{O}}(N^2 t)$ in the interaction-picture method, this indicates that our approach becomes advantageous in the large-system regime (large $N$), while the interaction-picture method is more favorable in the long-time and high-precision regime (large $t$ and small $\epsilon$). More precisely, the crossover between the two scalings occurs when $N = \Tilde{\mathcal{O}}(t/\epsilon)$. For $N$ beyond this threshold, our method achieves a lower overall gate complexity, whereas below this regime, the interaction-picture approach can be more efficient. While Ref.~\cite{low2018interactionpicture} does not provide an analysis for circuit depth, we assume perfect parallelism in their method, i.e., their circuit depth scales as $\mathcal{O}(N_G/N)$. This leads to a comparison that is qualitatively the same as that based on gate complexity. Moreover, our method has a lower ancilla scaling with system size $N$. It has been shown that our method requires $N_{\text{ancilla}}$ ancilla qubits, which scales linearly with $N$, as shown in Eq.~\eqref{eq:n_anci}. In contrast, a natural implementation of the interaction-picture method requires $\mathcal{O}\!\left(N(\log N+\log(1/\epsilon_b))\right)$ ancilla qubits to store $N$ binary numbers, each with bit length $\mathcal{O}(\log N+\log(1/\epsilon_b))$, when executing the fast Fourier transform; this number of ancilla qubits can be further simplified to $\Tilde{\mathcal{O}}(N)$.


Second, the main discussion in this work focuses on the extended Hubbard model with \emph{open} boundary conditions (OBC), while the work by G.~H.~Low and N.~Wiebe~\cite{low2018interactionpicture} discussed \emph{periodic} boundary conditions (PBC). This indicates that the two methods are designed for different settings. However, we also noticed that Q2FMM can be naturally extended to periodic boundary conditions (PBC) defined on a finite torus by constructing the interaction lists and redefining box-box displacement vectors using the wrapped (torus) distance. Analogous treatments have already been used in classical FMM implementations with the same asymptotic scaling and only minor additional cost~\cite{schwichtenberg1999acceleration, yokota2013petascale}.
Meanwhile, the interaction-picture-based method~\cite{low2018interactionpicture} could also be extended to OBC settings by enlarging the system and employing zero-padding~\cite{caprace2021flups}. Therefore, the distinction in boundary conditions should be viewed primarily as a difference in the target setting of the two works, rather than as a fundamental limitation of either method.

Therefore, a natural application where Q2FMM can offer an advantage is short- to intermediate-time dynamics simulation, such as correlation function measurement~\cite{PhysRevA.111.062610, PhysRevResearch.2.033281, kokcu2024linear}, estimating dynamical structure factor~\cite{baez2020dynamical}, or real-time dynamics following a global quench~\cite{PRXQuantum.2.010342}. One might expect that local observables could be estimated classically using a light-cone argument~\cite{kliesch2014lieb}. However, this reasoning applies primarily to local Hamiltonians. In the extended Hubbard model considered here, the interaction is long-ranged and all-to-all, and thus such an approximation may not directly apply.


\subsection{Optional optimizations for improving efficiency}

Here, we introduce two techniques, namely the \texttt{COPY} operation and the unbounded fan-out gate. These techniques are not the core novelty of Q2FMM, but can further improve the efficiency of particular circuit realizations.

Since the terms in the Coulomb interaction commute, the corresponding phases can be evaluated simultaneously for each interaction term within a given level. Thus, an efficient strategy for computing the phases is to first \texttt{COPY} the multipole expansions and then compute the evolving phase in parallel. The \texttt{COPY} operation, also commonly known as \emph{Transversal CNOT} \cite{Nielsen_2011, gottesman2009, Sahay_2025} or \emph{Element-wise CNOT}~\cite{dellachiara2025lcu}, is exemplified in Fig.~\ref{fig:copy_cnot}. Concretely, in our case, we need to \texttt{COPY} the multipole expansion of each box $\abs{I(A)}$ times, where $\abs{I(A)} \le 27$ due to the geometric setup. Thus, the number of required ancilla qubits scales linearly with the number of boxes.
With these ``copies,'' the phase evolution for all interacting box pairs can be computed in parallel, after which the \texttt{COPY} operation is uncomputed to disentangle the ancilla registers. Therefore, the quantum circuit at each level effectively requires only a single round of time-evolution computation, rather than $\abs{I(A)}$ sequential layers. Even though the asymptotic scaling of depth is not affected, the \texttt{COPY} operation significantly reduces the circuit depth without modifying the underlying algorithmic structure.

\begin{figure}[b!] 
  \centering
  \begin{quantikz}[row sep=0.1cm]
\lstick{$\ket{c_1}$} & \ctrl{2} & \qw      &  \ghost{\targ{}}\\
\lstick{$\ket{c_2}$} & \qw      & \ctrl{2} &  \ghost{\targ{}}\\
\lstick{$\ket{0}$} & \targ{}  & \qw      &  \ghost{\targ{}}\\
\lstick{$\ket{0}$} & \qw      & \targ{}  & \ghost{\targ{}}
\end{quantikz}
  \caption{``Copying'' occupation numbers to ancilla qubits via element-wise CNOTs, exemplified by a two-qubit state and two ancilla qubits. }
  \label{fig:copy_cnot}
\end{figure}

Another possible tool for reducing the depth of specific subroutines is the unbounded fan-out gate~\cite{hoyer2005quantum}, in which a single qubit simultaneously controls multiple target qubits, as illustrated in Fig.~\ref{fig:fanout}. Here, ``unbounded'' means we treat the fan-out operation as a constant-depth primitive, independent of the number of targets. Recent theoretical and experimental progress demonstrated the realization of the fan-out gate in shallow depth~\cite{khazali2020fast, young2021asymmetric, song2025constantdepth, baumermeasure, edp2022Beverland}, enabling further optimizations of our quantum algorithm. 

\begin{figure}[b!]
    \centering
    \begin{quantikz}
& \ctrl{3}     & \qw \\
& \targ{}       & \qw \\
& \targ{}        & \qw \\
& \targ{}       & \qw
\end{quantikz}
=\begin{quantikz}
& \ctrl{1}     & \ctrl{2} & \ctrl{3} & \qw\\
 & \targ{}     & \qw      & \qw      & \qw\\
& \qw          & \targ{}  & \qw      & \qw\\
& \qw          & \qw      & \targ{}  & \qw
\end{quantikz}
    \caption{Definition of the fan-out gate illustrated with four qubits.}
    \label{fig:fanout}
\end{figure} 

It is worth noting that, without utilizing the FMM approximation, the unbounded fan-out gate can perform a single Trotter step for the long-range Coulomb term in depth $\mathcal{O}(1)$. As proven in~\cite{hoyer2005quantum}, the commuting gates can be applied to the same qubits simultaneously with unbounded fan-out gates. However, this procedure needs $\mathcal{O}(N)$ ancilla qubits per site and thus $\mathcal{O}(N^2)$ ancilla qubits in total. To mitigate the quadratic ancilla blow-up while preserving low depth, we can integrate the fan-out gate into Q2FMM, achieving a more favorable ancilla-depth trade-off. Specifically, we employ the unbounded fan-out gate in both the \texttt{COPY} operation and the arithmetic circuits, i.e., quantum adder and multiplier. For the \texttt{COPY} operation, the fan-out gate enables us to \texttt{COPY} the occupation information of each box for $\abs{I(A)}$ ``copies'' with depth $1$.

For arithmetic operations, P.~H\o yer and R.~\v{S}palek~\cite{hoyer2005quantum} proved that the depth of the arithmetic circuits can be reduced to $\mathcal{O}(\log^{\star} n)$ by employing the unbounded quantum fan-out gate, where $\log^{\star} n$ (the iterated logarithm) is the number of times the logarithm must be applied to $n$ to reduce it to at most $1$.  Here, $n$ is the operand bit length. The iterated logarithm function $\log^{\star}n$ is a sublogarithmic function that grows very slowly with $n$. Even for input sizes as large as the estimated number of atoms in the observable universe, $\log^{\star}n \le 5$. Consequently, $\log^{\star}n$ can be treated as a \emph{constant} for all practical input sizes. Therefore, the total circuit depth scales as the number of levels, i.e., $\mathcal{O}(\log N)$. As detailed in Table~\ref{tab:fanout_table}, incorporating it into Q2FMM yields a more favorable ancilla-depth trade-off than a direct fan-out-only implementation of the Coulomb step.

\begin{table}[htbp]
\centering
\caption{The circuit depth and overall ancilla qubits needed to carry out a single Trotter step for the Coulomb term using the unbounded fan-out gate.}
\label{tab:fanout_table}
\renewcommand{\arraystretch}{0.95}
\setlength{\tabcolsep}{6pt}
\small
\begin{tabular}{cccc c}
\hline\hline
Method & Circuit Depth & \# Ancillae\\
\hline
Fan-out only
  & $\mathcal{O}(1) $
  & $\mathcal{O}(N^2)$  \\
Q2FMM + Fan-out
  & $\mathcal{O}(\log N)$   
  & $\Tilde{\mathcal{O}}(N)$  \\
\hline\hline
\end{tabular}
\end{table}

\section{Discussion}

In this work, we present Q2FMM, a quantum algorithm with polylogarithmic single-step circuit depth for simulating the extended Hubbard model. The key idea is to leverage the hierarchical structure of the fast multipole method (FMM) to reformulate site-site interactions as interactions between coarse-grained boxes, as well as the reuse of aggregated information, thereby reducing the computational cost scaling of long-range Coulomb interactions. It is worth emphasizing that, although we describe Q2FMM within the FMM framework, the workflow is not strictly FMM. Rather, it is more appropriately viewed as a multipole expansion algorithm with a hierarchical structure, which is inspired by FMM. While the Q2FMM framework is general, we suggest that a 2D neutral-atom quantum computer~\cite{hollerith2022distance, tao2024highfidelity, wei2023braneparity, graham2022multi, evered2023high, bluvstein2022processor, schmid2024compiler, bluvstein2024logical, evered2025kitaev, manetsch2025tweezer, chiu2025continuous3000, xu2025neutralhubbard} that directly supports long-range entangling gates and atom shuttling~\cite{graham2022multi, evered2023high, bluvstein2022processor, schmid2024compiler, bluvstein2024logical}, or a surface-code setting on 2D lattice where long-range operations and fan-out gate~\cite{litinski2019game, edp2022Beverland} can be implemented in constant depth could be particularly suitable for implementing or accelerating Q2FMM.

Moreover, the FMM is applicable to a broad class of kernels, including Yukawa~\cite{greengard2002yukawa}, Helmholtz~\cite{cheng2006wideband, engquist2007directional, chew2001fastEM}, Stokes flow~\cite{tornberg2008stokes}, linear elasticity~\cite{liu2009fmbem, gimbutas2016mindlin}, polyharmonic~\cite{gumerov2006biharmonic}, and Riesz potentials~\cite{ying2004kifmm}.
Consequently, our Q2FMM implementation can potentially be extended to these models as well. In addition, Q2FMM is not limited to the extended Hubbard model; it is also potentially applicable to ab initio molecular Hamiltonians discretized on real-space grids, as discussed in Appendix~\ref{app:discretization}. The truncation of the power-law interaction (which decays with distance) can also be incorporated into Q2FMM. The number of levels can be reduced by terminating Q2FMM at a level where the distance between boxes exceeds the chosen truncation distance $\xi$. This yields a total number of levels $\mathcal{O}(\log \xi)$ to be executed.

We are aware of a very recent work by D.~W.~Berry et al.~\cite{berry2025qfmm}, which also employs the FMM for quantum simulation, focusing on molecular Hamiltonians. Beyond the difference in target problems, the two works differ fundamentally in the type of quantum data to which the FMM is applied. In Berry et al.~\cite{berry2025qfmm}, the quantum state is represented in first quantization, and consequently, a key algorithmic challenge is to sort or assign electrons into spatial boxes based on their positions, which are stored in quantum registers. In other words, the algorithm must perform data access controlled by position information that is itself quantum data~\cite{berry2025qfmm}. In our work, by contrast, the quantum state is represented by the occupation numbers of lattice sites; each qubit corresponds directly to a grid point on the lattice and is therefore already arranged according to its spatial position before the quantum circuit is executed. Consequently, the main quantum subroutines are also different: Berry et al. focus on quantum data movement and access for electron registers, whereas this issue does not arise in our setting, since the qubits encoding the occupation numbers are already naturally ordered by their positions on the two-dimensional lattice. Therefore, our method instead centers on the efficient reuse of intermediate results while avoiding garbage entanglement, achieved through hierarchical aggregation and subsequent uncomputing.
It would be of particular interest to extend their framework to the 2D extended Hubbard model and compare its performance directly with that of Q2FMM.

\begin{acknowledgments}
We thank Dominic W.~Berry for pointing out Ref.~\cite{low2018interactionpicture} and helpful discussions. This research is part of the Munich Quantum Valley, which is supported by the Bavarian State Government with funds from the Hightech Agenda Bayern Plus. We also thank the Munich Center for Quantum Science and Technology. The packages \texttt{quantikz}~\cite{kay2018quantikz} and \texttt{yquant}~\cite{desef2020yquant} were used for drawing the quantum circuit diagrams.
\end{acknowledgments}

\appendix

\section{Generalization to the spinful Fermi-Hubbard model}
\label{app:reconsider}

Restoring the spin allows each site to host up to two fermions (one per spin), and only slightly modifies the long-range term:
\begin{equation}
V_\text{C} = \sum_{a\ne b} \frac{1}{2}V_{ab}\,\hat{n}_{a}\hat{n}_{b},
\end{equation}
where we rewrite the number operators as
\begin{equation}
\hat{n}_{a}= \sum_{\sigma\in\{\uparrow,\downarrow\}} \hat{n}_{a\sigma},\qquad
\hat{n}_{a\sigma}= \hat{c}^\dagger_{a\sigma}\hat{c}_{a\sigma}.    
\end{equation}
Adapted to Q2FMM, each site $i$ is now encoded by \emph{two} adjacent physical qubits. The only change to the circuit appears at the second finest level $L=\log_4(N)-1$, where each site's information is propagated to their parent boxes. Here, we simply apply the method illustrated in Fig.~\ref{fig:rq_circuit} to account for the contributions from the \emph{two} qubits representing a site. Consequently, at the finest level, there are eight qubits (corresponding to four sites) that are merged into their parent box. Accordingly, we require one additional ancilla qubit to store the information for each box.

\section{Overall quantum circuit exemplified by a one-dimensional lattice}
\label{app:overall_circuit}

\begin{figure}[b!]
    \centering
    \resizebox{\columnwidth}{!}{%
        \usetikzlibrary{decorations.pathreplacing}

\begin{tikzpicture}[scale=0.4, >=stealth]

\tikzset{qubit/.style={
    shade, circle,
    ball color=blue!50!white,
    shading angle=45,
    minimum size=3mm,
    inner sep=0pt, outer sep=0pt}}

\foreach \i in {0,...,15}{
  \node[qubit] (q\i) at (2*\i,0) {};
}

\draw[thick, rounded corners] 
  ($(q0)+(-0.8,-1.2)$) rectangle ($(q3)+(0.8,1.2)$);

\draw[thick, rounded corners] 
  ($(q4)+(-0.8,-1.2)$) rectangle ($(q7)+(0.8,1.2)$);

\draw[thick, rounded corners] 
  ($(q8)+(-0.8,-1.2)$) rectangle ($(q11)+(0.8,1.2)$);

\draw[thick, rounded corners] 
  ($(q12)+(-0.8,-1.2)$) rectangle ($(q15)+(0.8,1.2)$);

\draw[very thick, rounded corners, dashed] 
  ($(q0)+(-0.6,-0.8)$)  rectangle ($(q1)+(0.6,0.8)$);

\draw[very thick, rounded corners, dashed] 
  ($(q2)+(-0.6,-0.8)$)  rectangle ($(q3)+(0.6,0.8)$);

\draw[very thick, rounded corners, dashed] 
  ($(q4)+(-0.6,-0.8)$)  rectangle ($(q5)+(0.6,0.8)$);

\draw[very thick, rounded corners, dashed] 
  ($(q6)+(-0.6,-0.8)$) rectangle ($(q7)+(0.6,0.8)$);

\draw[very thick, rounded corners, dashed,] 
  ($(q8)+(-0.6,-0.8)$)  rectangle ($(q9)+(0.6,0.8)$);

\draw[very thick, rounded corners, dashed] 
  ($(q10)+(-0.6,-0.8)$)  rectangle ($(q11)+(0.6,0.8)$);

\draw[very thick, rounded corners, dashed] 
  ($(q12)+(-0.6,-0.8)$)  rectangle ($(q13)+(0.6,0.8)$);

\draw[very thick, rounded corners, dashed] 
  ($(q14)+(-0.6,-0.8)$) rectangle ($(q15)+(0.6,0.8)$);

\end{tikzpicture}
    }
    \caption{One-dimensional lattice geometry and hierarchy of boxes used for the circuit instantiation in Fig.~\ref{fig:overall_circuit}.}
    \label{fig:1d_lattice}
\end{figure}
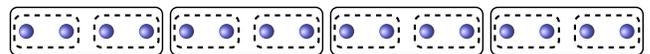

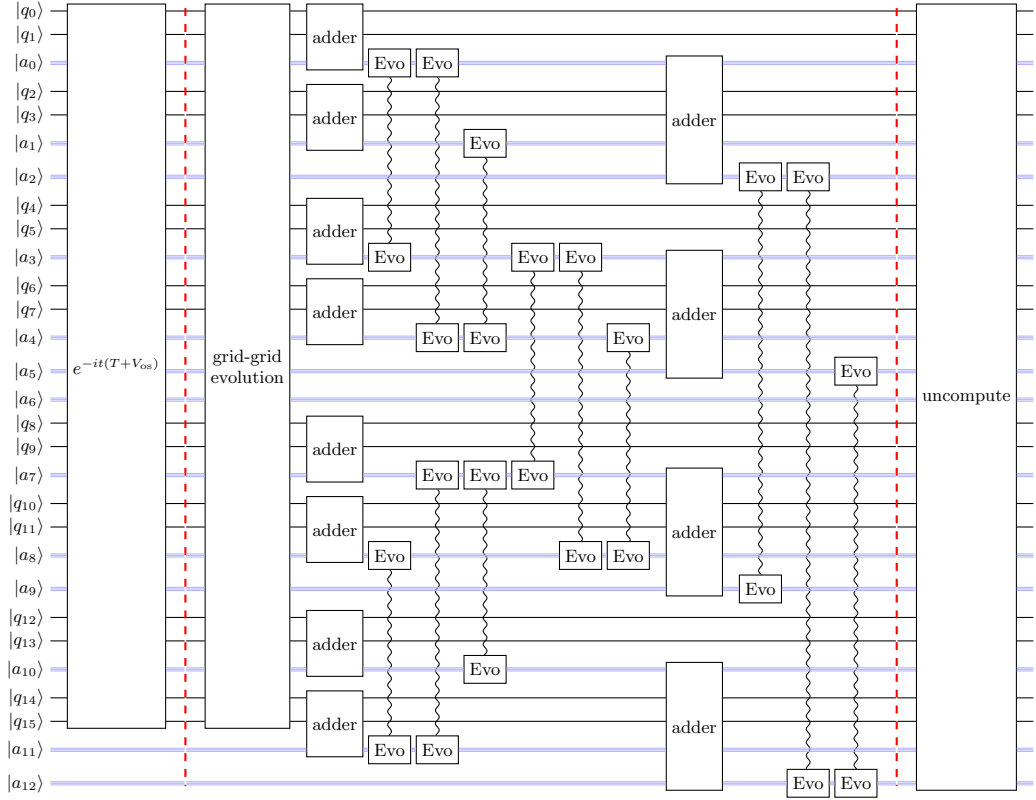
\begin{figure*}
    \centering
    \begin{tikzpicture} [scale = 0.75]
\begin{yquant} [operators/every barrier/.append style={red, thick}]
qubit {$\ket{q_0}$} a;  
qubit {$\ket{q_1}$} b;  
qubits {$\ket{a_0}$} c;  
qubit {$\ket{q_2}$} d;  
qubit {$\ket{q_3}$} e;  
qubits {$\ket{a_1}$} f;  
qubits {$\ket{a_2}$} g; 
qubit {$\ket{q_4}$} h;  
qubit {$\ket{q_5}$} i;  
qubits {$\ket{a_3}$} j; 
qubit {$\ket{q_6}$} k;  
qubit {$\ket{q_7}$} l; 
qubits {$\ket{a_4}$} m; 
qubits {$\ket{a_5}$} n; 
qubits {$\ket{a_6}$} o; 

qubit {$\ket{q_8}$} p;  
qubit {$\ket{q_9}$} q;  
qubits {$\ket{a_7}$} r;  
qubit {$\ket{q_{10}}$} s;  
qubit {$\ket{q_{11}}$} t;  
qubits {$\ket{a_8}$} u;  
qubits {$\ket{a_{9}}$} v; 
qubit {$\ket{q_{12}}$} w;  
qubit {$\ket{q_{13}}$} x;  
qubits {$\ket{a_{10}}$} y; 
qubit {$\ket{q_{14}}$} z;  
qubit {$\ket{q_{15}}$} 1; 
qubits {$\ket{a_{11}}$} 2; 
qubits {$\ket{a_{12}}$} 3; 

setstyle {blue!30} c;
setstyle {blue!30} f;
setstyle {blue!30} g;
setstyle {blue!30} j;
setstyle {blue!30} m;
setstyle {blue!30} n;
setstyle {blue!30} o;
setstyle {blue!30} r;
setstyle {blue!30} u;
setstyle {blue!30} v;
setstyle {blue!30} y;
setstyle {blue!30} 2;
setstyle {blue!30} 3;

hspace {2mm} -;

box {$e^{-i t (T+V_\text{os})}$}(a, b, c, d, e,f,g,h,i,j,k,l,m,n,o,p,q,r,s,t,u,v,w,x,y,z,1);

barrier (-);

box {grid-grid \\ evolution}(a, b, c, d, e,f,g,h,i,j,k,l,m,n,o,p,q,r,s,t,u,v,w,x,y,z,1);

hspace {2mm} -;
box {adder} (a, b, c);
box {adder} (d, e, f);
box {adder} (h, i, j);
box {adder} (k, l, m);
box {adder} (p, q, r);
box {adder} (s, t, u);
box {adder} (w, x, y);
box {adder} (z, 1, 2);

box {$\text{Evo}$} (c,j);
box {$\text{Evo}$} (c,m);
box {$\text{Evo}$} (u,2);
box {$\text{Evo}$} (r,2);
box {$\text{Evo}$} (y,r);
box {$\text{Evo}$} (f,m);
box {$\text{Evo}$} (j,r);
box {$\text{Evo}$} (j,u);
box {$\text{Evo}$} (m,u);

hspace {2mm} -;
align -;

box {adder} (c,d,e,f,g);
box {adder} (j,k,l,m,n);
box {adder} (r,s,t,u,v);
box {adder} (y,z,1,2,3);
hspace {2mm} -;
box {$\text{Evo}$} (g,v);
box {$\text{Evo}$} (g,3);
box {$\text{Evo}$} (n,3);

barrier (-);
box {uncompute}(a, b, c, d, e,f,g,h,i,j,k,l,m,n,o,p,q,r,s,t,u,v,w,x,y,z,1,2,3);

hspace {2mm} -;

\end{yquant} 
\end{tikzpicture}
    \caption{The overall quantum circuit implementing the $0^\text{th}$-order Q2FMM, instantiated for a 1D lattice (see Fig.~\ref{fig:1d_lattice}) for visual clarity. The qubits corresponding to physical sites are labeled by $\ket{q_i}$, while the ancilla registers are labeled by $\ket{a_i}$ and represented by blue bundle lines. The kinetic and on-site time evolution operator $e^{-i t (T+V_\text{os})}$ and the so-called grid-grid evolution only act on physical sites but not on the ancilla registers. The \texttt{Evo} gates acting on different boxes can be executed in parallel (but are displayed sequentially in this circuit diagram).}
    \label{fig:overall_circuit}
\end{figure*}

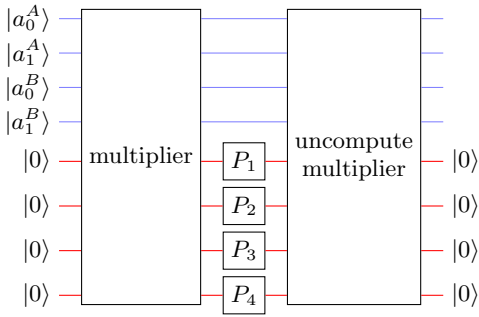
\begin{figure}
    \centering
    \begin{tikzpicture}

\begin{yquant}

qubit {$\ket{a^A_0}$} a;  
qubit {$\ket{a^A_1}$} b;  
qubit {$\ket{a^B_0}$} c;  
qubit {$\ket{a^B_1}$} d;  

qubit {$\ket{0}$} e;  
qubit {$\ket{0}$} f;  
qubit {$\ket{0}$} g;  
qubit {$\ket{0}$} h;  

setstyle {blue!50} a;
setstyle {blue!50} b;
setstyle {blue!50} c;
setstyle {blue!50} d;
setstyle {red} e;
setstyle {red} f;
setstyle {red} g;
setstyle {red} h;

hspace {2mm} -;

box {multiplier} (a, b, c, d, e, f, g, h);

hspace {2mm} -;

box {$P_1$} (e);
box {$P_2$} (f);
box {$P_3$} (g);
box {$P_4$} (h);

hspace {2mm} -;

box {uncompute\\multiplier} (a, b, c, d, e, f, g, h);

hspace {2mm} -;

output {$\ket{0}$} e;
output {$\ket{0}$} f;
output {$\ket{0}$} g;
output {$\ket{0}$} h;

\end{yquant}

\end{tikzpicture}
    \caption{An example of the \texttt{Evo} gate applied to the ancilla registers $\ket{a^A}$ and $\ket{a^B}$ corresponding to boxes $A$ and $B$, respectively. The additional ancilla qubits (red) are immediately discarded in a 2D neutral atom quantum computer after use, since uncomputing the quantum multiplier resets them to the $\ket{0}$ state. For this reason, they are not drawn in Fig.~\ref{fig:overall_circuit}.}
    \label{fig:evo_gate}
\end{figure}

We illustrate the structure of the $0^\text{th}$-order Q2FMM in Fig.~\ref{fig:overall_circuit}. For visual clarity, the circuit corresponds to a one-dimensional (1D) lattice with 16 sites, as shown in Fig.~\ref{fig:1d_lattice}, and ignores spin degrees of freedom. The overall quantum circuit for a single Trotter step is composed of three main components: the time evolution $e^{-i \delta t (T+V_\text{os})}$ for the hopping term $T$ and the on-site term $V_{\text{os}}$, the compute phase, and the uncompute phase, which are separated by red slices in Fig.~\ref{fig:overall_circuit}. A detailed description of the compute and uncompute phases, which aim to evaluate the time evolution governed by the long-range Coulomb term, is provided in Sec.~\ref{sec:0thcircuit}.

The compute phase begins by evaluating the time evolution at the finest level, where only near-neighbor interactions are involved and can be implemented straightforwardly; hence, we do not detail it here. When reaching a coarser level, we employ quantum adders to ``merge'' the boxes (grids at the finest level), as illustrated in Sec.~\ref{sec:0th_cost} and Fig.~\ref{fig:point1} of the main text. The blue bundle lines denote the ancilla registers, which are used to store the results of quantum adders, i.e., the summation of occupation numbers. The time evolution between boxes at a given level is then evaluated by the \texttt{Evo} gate, which is displayed as non-adjacent multi-qubit gates connected by a wave line. We omit the non-participating qubits between the two halves of the \texttt{Evo} gate and place the participating ancilla registers next to each other to illustrate the \texttt{Evo} gate in Fig.~\ref{fig:evo_gate}. We first employ a quantum multiplier, where extra ancilla qubits (highlighted in red) are introduced to record the result, to compute the product $N_A N_B$ of two boxes $A$ and $B$. Then, we use the phase gates in Eq.~\eqref{eq:phase} to generate the phases, and finally uncompute the multiplier, which is realized by its inverse gate. All pairs of ancilla registers on which \texttt{Evo} is applied are determined by the interaction list of each box, as detailed in Sec.~\ref{sec:0thcircuit}. The compute phase terminates upon completing the level that contains four boxes on a 1D lattice.

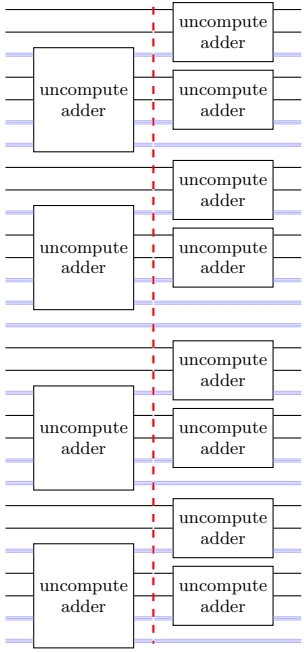
\begin{figure}
    \centering
    \begin{tikzpicture} [scale = 0.75]
\begin{yquant} [operators/every barrier/.append style={red, thick}]

qubit {} a ;  
qubit {} b;  
qubits {} c;  
qubit  {}d;  
qubit  {}e;  
qubits  {}f;  
qubits  {}g; 
qubit  {}h;  
qubit  {}i;  
qubits  {}j; 
qubit  {}k;  
qubit  {}l; 
qubits  {}m; 
qubits  {}n; 
qubits  {}o; 

qubit  {}p;  
qubit  {}q;  
qubits  {}r;  
qubit  {}s;  
qubit {}t;  
qubits  {}u;  
qubits  {}v; 
qubit  {}w;  
qubit  {}x;  
qubits {}y; 
qubit  {}z;  
qubit {}1; 
qubits  {}2; 
qubits {}3;

setstyle {blue!30} c;
setstyle {blue!30} f;
setstyle {blue!30} g;
setstyle {blue!30} j;
setstyle {blue!30} m;
setstyle {blue!30} n;
setstyle {blue!30} o;
setstyle {blue!30} r;
setstyle {blue!30} u;
setstyle {blue!30} v;
setstyle {blue!30} y;
setstyle {blue!30} 2;
setstyle {blue!30} 3;

hspace {4mm} -;


box {uncompute\\adder} (c,d,e,f,g);
box {uncompute\\adder} (j,k,l,m,n);
box {uncompute\\adder} (r,s,t,u,v);
box {uncompute\\adder} (y,z,1,2,3);

barrier (-);

box {uncompute\\adder} (a, b, c);
box {uncompute\\adder} (d, e, f);
box {uncompute\\adder} (h, i, j);
box {uncompute\\adder} (k, l, m);
box {uncompute\\adder} (p, q, r);
box {uncompute\\adder} (s, t, u);
box {uncompute\\adder} (w, x, y);
box {uncompute\\adder} (z, 1, 2);

hspace {4mm} -;

\end{yquant} 
\end{tikzpicture}
    \caption{Applied in reverse order, the uncompute procedure first ``splits'' the largest boxes, as also illustrated in Fig.~\ref{fig:box_uncompute}, and continues until the finest level is reached.}
    \label{fig:uncompute}
\end{figure}

The final step is to execute the uncompute phase, which transfers the phase back to the qubits representing the quantum state. As discussed in Sec.~\ref{sec:0thcircuit}, this is achieved by applying the inverses of the quantum arithmetic circuits in the reverse order. Since the multipliers have already been uncomputed in this example, it suffices to apply the inverses of the quantum adders, as shown in Fig.~\ref{fig:uncompute}.

\section{Trotter error of the 2D extended Hubbard model based on Commutators}
\label{app:commutator}

The theory developed in Ref.~\cite{Childs2021PRXCommutator} provides a tighter error bound for time evolution based on Trotterization than the direct norm-summation bound based on $\sum_i \norm{H_i}$, where $H = \sum_i H_i$, and $\norm{\cdot}$ denotes spectral norm. Here, we estimate the Trotter error for simulating the 2D extended Hubbard model.

We define
\begin{equation}
    V = V_{\text{os}} + V_\text{C}
\end{equation}
where $V_{\text{os}}$ and $V_\text{C}$ (see Eq.~\eqref{eq:hubbard}) commute with each other.
Hence, $H = T + V$ and the time evolution for a single time step $\delta t$ can be approximated as:
\begin{equation}
    \begin{aligned}
        U(\delta t) &= e^{-i \delta t (T+V)} \\
        & \approx e^{-i \delta t T/2} e^{-i \delta t V}  e^{-i \delta t T/2} + \mathcal{O}(\Lambda \, \delta t ^3),
    \end{aligned}
\end{equation}
by the second-order Trotterization, which leads to Eq.~\eqref{eq:trotter} in the main text. Here, $\Lambda$ can be estimated as:
\begin{equation}
\label{eq:trotter_error_theory}
    \Lambda \le \frac{1}{12}\|[V,[V,T]]\| + \frac{1}{24}\|[T,[T,V]]\|
\end{equation}
as discussed in Ref.~\cite{Childs2021PRXCommutator, ansgar2023hubbard}. Therefore,  the core task is to estimate $\|[V,[V,T]]\|$ and $\|[T,[T,V]]\|$. These two terms share a common contribution:
\begin{equation}
    \|[V,T]\| \le \|[V_{\text{os}},T]\| + \|[V_\text{C},T]\|.
\end{equation}
For convenience, we define $T_{ab, \sigma} = \hat{c}_{a, \sigma}^\dagger \hat{c}_{b, \sigma}$, then $\|[V_{\text{os}},T]\|$ can be expressed as:
\begin{equation}
    [V_{\text{os}}, T] = \sum_{\substack{\langle a,b \rangle\\ \sigma}} [V_{\text{os}}, T_{ab, \sigma}].
\end{equation}
Evaluating the commutator gives:
\begin{equation}
\begin{aligned}
[V_{\mathrm{os}}, T_{ab,\sigma}]
&=
V_0 \sum_i
[\hat n_{i\uparrow}\hat n_{i\downarrow}, \hat c^\dagger_{a\sigma}\hat c_{b\sigma}] \\
&=
V_0 \sum_i \hat n_{i\sigma'}[\hat n_{i\sigma}, \hat c^\dagger_{a\sigma}\hat c_{b\sigma}] \\
&=
V_0 \sum_i \hat n_{i\sigma'}([\hat n_{i\sigma}, \hat c^\dagger_{a\sigma}]\hat c_{b\sigma} + \hat c^\dagger_{a\sigma}[\hat n_{i\sigma}, \hat c_{b\sigma}])\\
&=
V_0 \sum_i \hat n_{i\sigma'}(\delta_{ia}  \hat c^\dagger_{a\sigma} \hat c_{b\sigma} - \delta_{ib} \hat c^\dagger_{a\sigma} \hat c_{b\sigma})\\
&=
V_0 T_{ab, \sigma}(\hat n_{a, \sigma ^{\prime}} - \hat n_{b, \sigma ^{\prime}}),
\end{aligned}
\end{equation}
so that 
\begin{equation}
\label{eq:vos_t}
    [V_{\text{os}}, T] = \sum_{\substack{\langle a,b \rangle\\ \sigma}} V_0 T_{ab, \sigma}(\hat n_{a, \sigma ^{\prime}} - \hat n_{b, \sigma ^{\prime}}),
\end{equation}
with $\sigma \ne \sigma^{\prime}$. Alternatively, the derivation can also be implemented by the physical meaning of $[V_{\text{os}},T]$, since 
\begin{equation}
    [V_{\text{os}},T_{ab,
    \sigma}] = V_{\text{os}}T_{ab,\sigma} - T_{ab,\sigma}V_{\text{os}}
\end{equation}
can be interpreted by the on-site potential difference caused by hopping. The on-site potential for sites $a$ and $b$ before hopping can be described by:
\begin{equation}
\hat n_{b,\sigma}\hat n_{b,\sigma^{\prime}}+\hat n_{a,\sigma}\hat n_{a,\sigma^{\prime}}.
\end{equation}
When considering $N_{b,\sigma}=1$ and $N_{a,\sigma}=0$, this can be simplified to:
\begin{equation}
    \hat n_{b,\sigma}\hat n_{b,\sigma^{\prime}}+\hat n_{a,\sigma}\hat n_{a,\sigma^{\prime}} = n_{b,\sigma^{\prime}}.
\end{equation}
Similarly, the one after hopping is:
\begin{equation}
\hat n_{b,\sigma}\hat n_{b,\sigma^{\prime}}+\hat n_{a,\sigma}\hat n_{a,\sigma^{\prime}} = \hat n_{a,\sigma^{\prime}},
\end{equation}
considering $N_{b,\sigma}=0$ and $N_{a,\sigma}=1$. Therefore, we immediately reach Eq.~\eqref{eq:vos_t}.

Next, we estimate 
\begin{equation}
    [V_\text{C}, T] =  \sum_{\substack{\langle a,b \rangle\\ \sigma}}[V_\text{C}, T_{ab,\sigma}].
\end{equation}
Observe the effect of applying $[V_\text{C}, T_{ab,\sigma}]$ to a computational basis state $\ket{n}$:
\begin{equation}
    [V_\text{C}, T_{ab,\sigma}] \ket{n}= V_\text{C} T_{ab,\sigma} \ket{n} - T_{ab,\sigma} V_\text{C} \ket{n},
\end{equation}
$V_\text{C} T_{ab,\sigma} \ket{n}$ moves the electron from $b$ to $a$ and then evaluates the total Coulomb potential; the second term $T_{ab,\sigma} V_\text{C} \ket{n}$ evaluate the total Coulomb interaction, and then move the electron from $b$ to $a$. Note that $V_\text{C} T_{ab,\sigma} \ket{n}$ and $T_{ab,\sigma} V_\text{C} \ket{n}$ finally reach the same state while having different coefficients corresponding to different energies. Therefore, $[V_\text{C}, T_{ab,\sigma}] \ket{n}$ has a clear physical meaning, which is evaluating the difference of the total Coulomb interaction caused by the hopping. Thus, $[V_\text{C}, T_{ab,\sigma}]$ can also be expressed as:
\begin{equation}
\begin{aligned}
    &[V_\text{C}, T_{ab,\sigma}] \\
    =& T_{ab,\sigma}
    \left[\sum_{c\ne a,b}(V_{ac}-V_{bc})(\hat n_{c,\uparrow} + \hat n_{c,\downarrow})+V_{ab}\bigl(\hat n_{b,\sigma^{\prime}}-\hat n_{a,\sigma^{\prime}}\bigr)\right],
\end{aligned}
\end{equation}
remind that $\sigma^{\prime}$ denotes the spin opposite to $\sigma$,
Here, the first term describes the interaction change related to a grid point $c \ne a,b$, and the second term describes the energy change only related to $a$ and $b$.

In summary, 
\begin{equation}
\begin{aligned}
[V,T_{ab,\sigma}]
&=
[V_{\text{os}},T_{ab,\sigma}] + [V_C,T_{ab,\sigma}]
\\[6pt]
&=
T_{ab,\sigma}\Bigg[\sum_{c\ne a,b}(V_{ac} -V_{bc})
(\hat n_{c,\uparrow} +\hat n_{c,\downarrow})
\\
&\qquad\qquad
+ (V_{ab}-V_0)\bigl(\hat n_{b,\sigma'}-\hat n_{a,\sigma'}\bigr)\Bigg].
\end{aligned}
\end{equation}
For convenience, we define:
\begin{equation}
    \mathcal{F}_{ab,\sigma} = \sum_{c\ne a,b}(V_{ac}-V_{bc})
(\hat n_{c,\uparrow}+\hat n_{c,\downarrow})+
(V_{ab}-V_0)
\bigl(
\hat n_{b,\sigma'}-\hat n_{a,\sigma'}
\bigr),
\end{equation}
hence
\begin{equation}
    [V,T_{ab,\sigma}] = T_{ab,\sigma} \mathcal{F}_{ab,\sigma}.
\end{equation}
Further, we have:
\begin{equation}
\begin{aligned}
[V,[V,T_{ab,\sigma}]]
&=
[V,T_{ab,\sigma}\mathcal{F}_{ab,\sigma}]
\\
&=
[V,T_{ab,\sigma}]\,\mathcal{F}_{ab,\sigma}
+ T_{ab,\sigma}[V,\mathcal{F}_{ab,\sigma}] \\
&= [V, T_{ab,\sigma}] \mathcal{F}_{ab,\sigma}
\end{aligned}
\end{equation}
due to $[V, \mathcal{F}_{ab,\sigma}] = 0$, since $V$ and $\mathcal{F}_{ab,\sigma}$ only contain number operators and thus commute with each other. Hence:
\begin{equation}
    [V,[V,T_{ab,\sigma}]] = [V, T_{ab,\sigma}] \mathcal{F}_{ab,\sigma} = T_{ab,\sigma} \mathcal{F}^2_{ab,\sigma},
\end{equation}
and:
\begin{equation}
    [V,[V,T]] = \sum_{\substack{\langle a,b \rangle\\ \sigma}} T_{ab,\sigma} \mathcal{F}^2_{ab,\sigma}.
\end{equation}

We now compute $\|[V,[V,T]]\|$:
\begin{equation}
\begin{aligned}
    \| [V,[V,T]]\| &\le \sum_{\substack{\langle a,b \rangle\\ \sigma}} \| T_{ab,\sigma} \| \cdot \| \mathcal{F}^2_{ab,\sigma}\| \\
    &\le h\sum_{\substack{\langle a,b \rangle\\ \sigma}} \|   \mathcal{F}^2_{ab,\sigma}\| \\
    &\le h\sum_{\substack{\langle a,b \rangle\\ \sigma}}    \Bigg[\sum_{c\ne a,b}2|V_{ac}-V_{bc}| + |V_{ab} - V_0| \Bigg]^2,
\end{aligned}
\end{equation}
remember that $h$ is the hopping amplitude. Noting that the grid points $a$ and $b$ are neighbors, and assuming that the point $c$ is far away from $a$ and $b$, we obtain:
\begin{equation}
    V_{ac} - V_{bc} = \frac{1}{r_{ac}} - \frac{1}{r_{bc}} = \mathcal{O}\!\left(\frac{1}{r^2}\right),
\end{equation}
where $r \approx r_{ac} \approx r_{bc}$ is the distance between $c$ and $a,b$. There are $\mathcal{O}(r)$ grid points at distance $r$, and we define $A(r)$ as the collection of grid points at distance $r$, hence:
\begin{equation}
\begin{aligned}
    \sum_{c\ne a,b}2(V_{ac}-V_{bc}) &= \sum_{r=1}^{\xi_{\text{max}}} \sum_{c\in A(r)} 2(V_{ac}-V_{bc}) \\
    &= \sum_{r=1}^{\xi_{\text{max}}} \mathcal{O}(r) \cdot 1/r^2 \\
    &= \sum_{r=1}^{\xi_{\text{max}}} \mathcal{O} (1/r) \\
    &= \mathcal{O}(\log (\xi_{\text{max}}))\\
    &= \mathcal{O}(\log \sqrt{N}),
\end{aligned}
\end{equation}
where $\xi_{\text{max}}$ is the linear size of the lattice and $N$ is the total number of grid points. While $V_{ab} -V_0$ is constant, $\sum_{c\ne a,b}2(V_{ac}-V_{bc})$ is system-size dependent and therefore determines the asymptotic scaling. Thus:  
\begin{equation}
\label{eq:f_norm}
     \| \mathcal{F}_{ab,\sigma}\| \sim \mathcal{O}(\log N),
\end{equation}
and
\begin{equation}
\begin{aligned}
    \| [V,[V,T]]\| 
    &\le h\sum_{\substack{\langle a,b \rangle\\ \sigma}} \|   \mathcal{F}^2_{ab,\sigma}\| \\
    &\le h\sum_{\substack{\langle a,b \rangle\\ \sigma}} \mathcal{O}(\log^2 \sqrt{N})\\
    &\sim \mathcal{O}(N \log^2 N),
\end{aligned}
\end{equation}
where we suppress constant prefactors, such as the hopping amplitude and the on-site interaction strength, since our goal is to determine the asymptotic scaling with the system size $N$.
We will see that $\| [V,[V,T]]\|$ provides the leading contribution to the asymptotic scaling of $\Lambda$.

The remaining task is to estimate $\| [T,[T,V]]\|$, for which we can show that it is only secondary to $\| [V,[V,T]]\|$.
We can still use the conclusion that $[V,T_{ab,\sigma}] = T_{ab,\sigma} \mathcal{F}_{ab,\sigma}$, which leads to:
\begin{equation}
\label{eq:ttv}
\begin{aligned}
\|[T,[T,V]]\| &\le \|\sum_{\substack{\langle a,b \rangle\\ \sigma}} [T, T_{ab,\sigma} \mathcal{F}_{ab,\sigma}] \|\\
&= 
\|\sum_{\substack{\langle a,b \rangle\\ \sigma}} [T, T_{ab,\sigma}]\mathcal{F}_{ab,\sigma}\| +
\|\sum_{\substack{\langle a,b \rangle\\ \sigma}} T_{ab,\sigma} [T,  \mathcal{F}_{ab,\sigma}]\|.
\end{aligned}
\end{equation}
Observe the first term, and we can realize that $[T, T_{ab,\sigma}]$ is local, since only hopping terms that are related to sites $a$ and $b$ do not commute with $T_{ab,\sigma}$. Hence, we immediately have:
\begin{equation}
    \| [T, T_{ab,\sigma}]\| \sim \mathcal{O}(1),
\end{equation}
which is not related to the system size $N$. Hence, the norm of the first term in $[T,[T,V]]$ can be bounded by:
\begin{equation}
\label{eq:TTabFab}
\begin{aligned}
    \|\sum_{\substack{\langle a,b \rangle\\ \sigma}} [T, T_{ab,\sigma}]\mathcal{F}_{ab,\sigma}\| &\le \sum_{\substack{\langle a,b \rangle\\ \sigma}}\| [T, T_{ab,\sigma}]\| \cdot \| \mathcal{F}_{ab,\sigma}\|\\
    &\le \sum_{\substack{\langle a,b \rangle\\ \sigma}} \mathcal{O}(1) \cdot \mathcal{O}(\log N)\\
    &\le \mathcal{O}(N \log N),
\end{aligned}
\end{equation}
where we suppress constant prefactors as before, and the result $ \| \mathcal{F}_{ab,\sigma}\| \sim \mathcal{O}(\log N)$ in Eq.~\eqref{eq:f_norm} is used.

We now evaluate $\sum_{\substack{\langle a,b \rangle\\ \sigma}} T_{ab,\sigma} [T,  \mathcal{F}_{ab,\sigma}]$:
\begin{equation}
\begin{aligned}
    \sum_{\substack{\langle a,b \rangle\\ \sigma}} T_{ab,\sigma} [T,  \mathcal{F}_{ab,\sigma}] = \sum_{\substack{\langle a,b \rangle\\ \sigma}} T_{ab,\sigma} \Big[\sum_{\substack{\langle c,d \rangle\\ \tau}}T_{cd,\tau},  \mathcal{F}_{ab,\sigma}\Big].
\end{aligned}
\end{equation}
Recall that $[V,T_{ab,\sigma}] = T_{ab,\sigma} \mathcal{F}_{ab,\sigma}$ describes the potential difference with a hopping $T_{ab,\sigma}$; similarly, $[T_{cd,\tau},  \mathcal{F}_{ab,\sigma}]$ measures how this potential difference is modified by another hopping $T_{cd,\tau}$. Therefore, intuitively, such a term can only contribute a smaller part to $[V, T_{ab,\sigma}]$. More concretely, the potential difference before this hopping is related to $V_{ac}- V_{bc}$, and the one after hopping is $V_{ad}- V_{bd}$, as discussed before. If $(a,b)$ and $(c,d)$ are far from each other, we obtain:
\begin{equation}
    (V_{ad}- V_{bd}) - (V_{ac}- V_{bc}) \approx \frac{1}{r^3},
\end{equation}
which can be interpreted as a second discrete difference of the Coulomb potential, where $r \approx r_{ac}\approx r_{bc}\approx r_{ad}\approx r_{bd}$. Therefore,
\begin{equation}
    \|[T_{cd,\tau},  \mathcal{F}_{ab,\sigma}]\| = \mathcal{O}\left(\frac{1}{r^3}\right),
\end{equation}
which leads to:
\begin{equation}
\label{eq:TabTFab}
\begin{aligned}
    \|\sum_{\substack{\langle a,b \rangle\\ \sigma}} T_{ab,\sigma} [T,  \mathcal{F}_{ab,\sigma}]\| 
    &\le \sum_{\substack{\langle a,b \rangle\\ \sigma}} \|T_{ab,\sigma}\| \sum_{\substack{\langle c,d \rangle\\ \tau}}\|[T_{cd,\tau},  \mathcal{F}_{ab,\sigma}]\| \\
    &\le \sum_{\substack{\langle a,b \rangle\\ \sigma}} \|T_{ab,\sigma}\| \cdot \sum_{r=1}^{\xi_{\text{max}}} \mathcal{O}(r) \cdot \mathcal{O}(1/r^3) \\
    &\sim \mathcal{O}(N).
\end{aligned}
\end{equation}
Substituting this result together with Eq.~\eqref{eq:TTabFab} into Eq.~\eqref{eq:ttv}, we obtain:
\begin{equation}
    \|[T,[T,V]]\| \le \mathcal{O}(N \log N) + \mathcal{O}(N),
\end{equation}
which simplifies asymptotically to
\begin{equation}
    \|[T,[T,V]]\| \le \mathcal{O}(N \log N),
\end{equation}
since we focus on the asymptotic scaling with $N$. Recall that 
\begin{equation}
    \| [V,[V,T]]\| \le \mathcal{O}(N \log^2 N),
\end{equation}
this term asymptotically dominates $\|[T,[T,V]]\|$, we finally reach:
\begin{equation}
\label{eq:final_bound}
    \Lambda \le  \mathcal{O}(N \log^2 N)
\end{equation}
employing Eq.~\eqref{eq:trotter_error_theory}.

The previous method~\cite{berry2007efficient,low2019well} estimates the Trotter error using a tail bound of the Taylor expansion, leading to $\Lambda \le \norm{T} + \norm{V} = \mathcal{O}(N^{3/2})$. With the commutator-based theory, we derived a tighter bound on $\Lambda$ given in Eq.~\eqref{eq:final_bound}, which in turn allows us to estimate the required number of Trotter steps and provide a more careful analysis of the overall gate complexity in the main text.

\section{Bringing box information together: the shuttling operation}
\label{app:shuttle}

The qubits encoding the occupation information of each box (and also their copies when \texttt{COPY} is utilized) are typically located at its corresponding geometric location. To evaluate the long-range interaction between two spatially separated boxes, either (i) the hardware must support long-range quantum gates or (ii) only local gates are available, and the relevant information must be brought together. Below, we discuss the latter case in detail. 

Under the constraint of nearest-neighbor connectivity, the interaction is typically implemented through a series of SWAP gates that effectively ``move'' the ancilla qubits of the two boxes into proximity. The circuit depth associated with these SWAP operations scales linearly with the distance $r_{AB}$ between the boxes $A$ and $B$.

Alternatively, if hardware supports qubit shuttling, i.e., physically transporting qubits between regions, one replaces a SWAP chain of depth $\mathcal{O}(r_{AB})$ by only two shuttling operations, reducing the routing depth to $\mathcal{O}(1)$. Shuttling operations, or functionally equivalent long-range coupling, have been the subject of extensive theoretical proposals and experimental realizations in multiple quantum computing platforms, such as the neutral atom quantum computer~\cite{graham2022multi,evered2023high,bluvstein2022processor,bluvstein2024logical,schmid2024compiler} and the trapped-ion quantum computer~\cite{kielpinski2002architecture, blakestad2009xjunction, schoenberger2025shuttling, pino2021qccd, moses2023racetrack, guo2024siteresolved, li2023kzm_longrange_tfim}.

\section{Grid discretization for ab initio molecular Hamiltonians}
\label{app:discretization}


Here, we illustrate how to apply a grid discretization to the following real-space representation of a molecular Hamiltonian (within the Born-Oppenheimer approximation):
\begin{equation}
\begin{split}
H = T + V = \sum_{\sigma} \int d\mathbf{r} \, \hat{\psi}^\dagger_{\mathbf{r} \sigma} \left( -\frac{1}{2} \nabla^2 + v(\mathbf{r}) \right) \hat{\psi}_{\mathbf{r} \sigma} \\
+ \frac{1}{2} \sum_{\sigma, \sigma'} \int d\mathbf{r} d\mathbf{r'} \,  \hat{\psi}^\dagger_{\mathbf{r} \sigma} \hat{\psi}^\dagger_{\mathbf{r'} \sigma'}  \frac{1}{\|\mathbf{r} - \mathbf{r'}\|_2 } \hat{\psi}_{\mathbf{r'} \sigma'} \hat{\psi}_{\mathbf{r} \sigma},
\end{split}
\label{eq:continuum_H}
\end{equation}
where $\hat{\psi}_{\mathbf{r}\sigma}$ and $\hat{\psi}_{\mathbf{r}\sigma}^{\dagger}$ are annihilation and creation operators for spin $\sigma$ at position $\mathbf{r}$, respectively; $v(\mathbf r)$ denotes the electron-nuclear potential. Employing the grid approximation, we replace the operator $\hat{\psi}_{\mathbf{r}\sigma}$ with grid annihilation operators:
\begin{equation}
    \hat{\psi}_{\mathbf{r} \sigma} \rightarrow \sum_{i=1}^N \delta(\mathbf{r} - \mathbf{r}_i) \hat{a}_{i \sigma}
\label{eq:grid_op}
\end{equation}
where $\hat{a}_{i \sigma}$ is the annihilation operator for grid point $i$, $r_i$ is the position of grid point $i$, and $N$ is the total number of grid points. Inserting Eq.~\eqref{eq:grid_op} into the Hamiltonian Eq.~\eqref{eq:continuum_H}, the Coulomb term $V$ can be re-written as:
\begin{equation}
\label{eq:grid_cou}
V = \frac{1}{2} \sum_{i \neq j} \frac{1}{\|\mathbf{r}_i - \mathbf{r}_j\|_2 }  \hat{n}_i \hat{n}_j,
\end{equation}
which only contains the grid-grid Coulomb interaction. The kinetic term $T$ only contains nearest-neighbor hopping terms by replacing $\nabla^2$ by finite-difference approximations.

The difficulty in simulating Eq.~\eqref{eq:grid_cou} using Q2FMM lies in the need for a three-dimensional grid setting to capture the electron's motion in three-dimensional space. An alternative is to map the three-dimensional space onto a two-dimensional one. A simpler case is plane- or chain-structured molecules, where electrons move freely in the first two dimensions but are confined to two dimensions in the third. In such cases, a two-dimensional lattice setting is sufficient.

\newpage
\bibliography{references}
\end{document}